\documentclass[aps,showpacs,floatfix,preprintnumbers,nofootinbib,11pt,prd,a4paper]{revtex4-2}%{revtex4-1}
\usepackage{graphicx}
\usepackage{dcolumn}
\usepackage{bm}
\usepackage{amsfonts}
\usepackage{amsthm}
\usepackage{amsmath}
\usepackage{amssymb}
\usepackage{epsfig}
\usepackage{colortbl}
\usepackage[usenames,dvipsnames]{xcolor}
\usepackage{tabularx}
\usepackage{tablefootnote}
\usepackage{textcomp}
\usepackage{hyperref}
\usepackage{titlesec}
\usepackage{slashed}
\usepackage{subcaption}
\usepackage[sort&compress]{natbib}

\newcommand{\nede}{_{_{\rm NEDE}}}
\newcommand{\kmsMpc}{\, \text{km}\,\text{s}^{-1}\, \text{Mpc}^{-1}}

\def\ee{\end{equation}}
\def\ba{\begin{eqnarray}}
\def\ea{\end{eqnarray}}

\def\bdm{\begin{displaymath}}
\def\edm{\end{displaymath}}
\numberwithin{equation}{section}

\newcommand{\beq}{\begin{equation}}
\newcommand{\eeq}{\end{equation}}
\newcommand{\beqa}{\begin{eqnarray}}
\newcommand{\eeqa}{\end{eqnarray}}
\newcommand{\bea}{\begin{eqnarray}}
\newcommand{\eea}{\end{eqnarray}}

\newcommand{\al}{\alpha}
 \newcommand{\be}{\beta}

\def\lesssim{~\mbox{\raisebox{-.6ex}{$\stackrel{<}{\sim}$}}~}

\def\ltap{\ \raise.3ex\hbox{$<$\kern-.75em\lower1ex\hbox{$\sim$}}\ }
\def\gtap{\ \raise.3ex\hbox{$>$\kern-.75em\lower1ex\hbox{$\sim$}}\ }
\def\gl{\ \raise.5ex\hbox{$>$}\kern-.8em\lower.5ex\hbox{$<$}\ }
\def\roughly#1{\raise.3ex\hbox{$#1$\kern-.75em\lower1ex\hbox{$\sim$}}}

\newcommand{\bi}{\begin{itemize}}
\newcommand{\ei}{\end{itemize}}

\def\ltap{\ \raise.3ex\hbox{$<$\kern-.75em\lower1ex\hbox{$\sim$}}\ }
\def\gtap{\ \raise.3ex\hbox{$>$\kern-.75em\lower1ex\hbox{$\sim$}}\ }
\def\gl{\ \raise.5ex\hbox{$>$}\kern-.8em\lower.5ex\hbox{$<$}\ }
\def\roughly#1{\raise.3ex\hbox{$#1$\kern-.75em\lower1ex\hbox{$\sim$}}}

\begin{document}

\title{Cold New Early Dark Energy pulls the trigger on the $H_0$ and $S_8$ tensions:\\ a simultaneous solution to both tensions without new ingredients}

\author{Juan S. Cruz\footnote{jcr@sdu.dk}${}^1$}\author{Florian Niedermann\footnote{florian.niedermann@su.se}$^{2}$}\author{Martin S.~Sloth\footnote{sloth@sdu.dk}${}^1$}

\affiliation{${}^1$ Universe Origins Group and CP$^3$-Origins\\ University of Southern Denmark, Campusvej 55, 5230 Odense M, Denmark}

\affiliation{${}^2$ Nordita, KTH Royal Institute of Technology and Stockholm University\\
Hannes Alfv\'ens v\"ag 12, SE-106 91 Stockholm, Sweden}


\begin{abstract}
In this work, we show that the Cold New Early Dark Energy (Cold NEDE) model in its original form can solve both the Hubble tension and the $S_8$ tension without adding any new ingredients at the fundamental level. So far, it was assumed that the trigger field in the Cold NEDE model is completely subdominant. However, relaxing this assumption and letting the trigger field contribute a mere $0.5\%$ of the total energy density leads to a resolution of the $S_8$ tension while simultaneously improving it as a solution to the $H_0$ tension.
Fitting this model to baryonic acoustic oscillations, large-scale-structure, supernovae (including a SH$_0$ES prior), and cosmic microwave background data, we report a preferred NEDE fraction of $f_\mathrm{NEDE}=  0.134^{+0.032}_{-0.025}$ ($68\%$ C.L.), lifting its Gaussian evidence for the first time above $5\sigma$ (up from $4 \sigma$ when the trigger contribution to dark matter is negligible). At the same time, we find the new concordance values $H_0 = 71.71 \pm 0.88 \,\mathrm{km}\, \mathrm{sec}^{-1}\, \mathrm{Mpc}^{-1}$ and $S_8 = 0.793 \pm 0.018$.  Excluding large-scale structure data and the SH$_0$ES prior, both Gaussian tensions are reduced below the $2 \sigma$ level.
\end{abstract}

\maketitle

\newpage

\tableofcontents

\section{Introduction}

The Hubble tension is a disagreement between the measurements of the expansion rate of the universe today, $H_0$, when measured using so-called early time measurements, such as the Cosmic Microwave Background (CMB) and Baryonic Acoustic Oscillation (BAO) and when using late-time (or local) measurements, such as supernovae or gravitational lensing time-delays (c.f. \cite{Aghanim2020, Riess2022, Beutler2011, Ross2015, Alam2017, Scolnic2018}). This tension is dependent on $\Lambda$CDM as our cosmological model and, by now, has been claimed to be $5 \sigma$ significant~\cite{Riess2022}. But in addition to the Hubble tension, a new similar tension is emerging. When measuring the amplitude of large-scale matter density perturbations through the $S_8$ parameter, a different result is obtained from a late-time (low-redshift) weak lensing measurement, such as weak lensing~\cite{Joudaki2020,DES:2021wwk,DES:2021vln,Busch:2022pcx,DES:2021wwk}, galaxy clustering~\cite{Zhai:2022yyk,Yuan:2022jqf}, and CMB lensing tomography~\cite{Krolewski:2021yqy,White:2021yvw,Chen:2022jzq,DES:2022urg}, than what is inferred from an early-time CMB measurement. The $\Lambda$CDM model predicts that matter clumps too much on small scales, which manifests itself as a larger value of $S_8$. This discrepancy has a significance of around 2-3 $\sigma$ (for reviews see~\cite{Abdalla2022,DiValentino:2020vvd}), although its robustness is still being debated (see, for example, the analysis in~\cite{DAmico:2022osl}).

It has been established that New Early Dark Energy (NEDE) is a natural and promising solution to the notorious Hubble tension~\cite{Niedermann:2019olb, Niedermann:2020dwg, Niedermann:2020qbw, Niedermann:2021ijp, Niedermann:2021vgd, Schoeneberg2022,Cruz:2023cxy,Cruz:2022oqk} and remains compatible with measurements of large-scale structure (LSS) \cite{Niedermann:2020qbw,Cruz:2023cxy} (for other adaptions of NEDE see also~\cite{Freese:2021rjq,Allali:2021azp}). So far, it has shared this phenomenological success with its predecessor Early Dark Energy (EDE)~\cite{Karwal:2016vyq, Poulin:2018cxd,Poulin:2018dzj,Smith:2019ihp,Smith:2022hwi}, although both models rely on different microphysics and lead to different predictions at the more detailed level (for further EDE-type models and explorations see~\cite{Lin:2019qug, Sakstein:2019fmf, Braglia:2020bym, Karwal:2021vpk, Freese:2021rjq, Allali:2021azp, Agrawal:2019lmo, Berghaus:2019cls, Berghaus:2022cwf, Ye:2020btb, Sabla:2022xzj, Gomez-Valent:2021cbe, Moss:2021obd, McDonough:2021pdg, Alexander:2022own, McDonough:2022pku, Gomez-Valent:2022bku} and for reviews~\cite{Kamionkowski:2022pkx,Poulin:2023lkg}). However, just like other EDE models, the NEDE model has been thought not to lead to a simultaneous alleviation of the $S_8$ problem. Since the $S_8$ problem is another early vs late universe tension, it has been argued that a natural solution to the Hubble tension should simultaneously address the $S_8$ tension. As we will show in this work, the NEDE model already holds in its original form a simultaneous resolution to the Hubble tension and the $S_8$ problem without the need to add any new ingredients. This is special for NEDE and sets it apart from other EDE models, which require additional new physics to be added independently to address the $S_8$ problem. In this way, NEDE now manifests itself as the most straightforward way out of a tense situation of cosmic proportions.

Unlike the old EDE model, NEDE is a \textit{fast-triggered} phase transition around matter-radiation equality before recombination. In the NEDE framework, different triggers of the phase transition are possible. So far, mostly the case of a scalar trigger, the Cold NEDE model~\cite{Niedermann:2019olb, Niedermann:2020dwg, Niedermann:2020qbw, Cruz:2022oqk, Cruz:2023cxy}, and the case of a dark sector temperature trigger, the Hot NEDE model \cite{Niedermann:2021ijp, Niedermann:2021vgd}, has been considered. In both cases, the trigger initiates a first-order phase transition, which leads to the decay of an eV scale dark energy component\footnote{A triggered second-order phase transition, hybrid NEDE, is also possible as briefly discussed in \cite{Niedermann:2020dwg}.}.

In the Cold NEDE model, the scalar trigger is an ultralight field, $\phi$ \cite{Niedermann:2019olb, Niedermann:2020dwg}.  Being very light compared to the Hubble rate, it is initially frozen in its potential due to the Hubble friction. However, as the Hubble rate decreases, the trigger field, $\phi$, becomes heavy compared to the Hubble scale just before matter-radiation equality. At this point, it becomes dynamic and triggers the phase transition (by lowering the potential barrier of another field $\psi$ initially stuck in a false minimum). That way, the Cold NEDE model predicts that the mass of the trigger field, $\phi$, is $m \approx 0.2 H_*$, where $H_*$ is the Hubble rate at the time of the phase transition. To solve the Hubble tension, the NEDE phase transition needs to happen at a redshift between $z_* = 4000 - 5000$, when $H_*$ takes a value of $H_* \approx 10^{-26}$ eV or equivalently $m \approx 10^{-27}$ eV.

Since the mass of the trigger field, $\phi$, is a prediction of the Cold NEDE model, only the initial value of the field is a free parameter. So far, it was assumed that the trigger field, $\phi$, is completely subdominant, which only gives an upper bound on its initial value, $\phi_\mathrm{ini}$, and it was therefore not included as a free parameter in the fit to the CMB, SN, and BAO data (even then the trigger leaves a characteristic fingerprint by setting the initial conditions in the decaying NEDE fluid, which is independent of $\phi_\mathrm{ini}$). Already in the original Cold NEDE papers, it was suggested that when relaxing this assumption of $\phi$ making a negligible contribution to the background energy density, it could lead to a resolution of the $S_8$ tension. However, at the time, it was still debated how seriously the $S_8$ tension should be taken.

In this work, we will take a step forward and consider the possibility that the trigger field, $\phi$, contributes a small fraction of the Dark Matter (DM) energy density. As this fraction is controlled by the initial $\phi$ value, it simply corresponds to relaxing the constraints on $\phi_\mathrm{ini}$ in the original Cold NEDE model without adding any new degrees of freedom.

Ultra-light scalar fields have been considered before in the literature as a source of dark matter~\cite{Preskill:1982cy,Turner:1983he,Brandenberger:1984jq,Ratra:1987rm,Kim:1986ax,Marsh:2015xka} and a possible solution to the $S_8$ problem~\cite{Hlozek:2017zzf,Lague:2021frh,Rogers:2023ezo}. Such Ultra-Light Axion-like particles (ULA) are theoretically well motivated and could arise as pseudo-Nambu-Goldstone bosons of spontaneously broken shift symmetries in string theory~\cite{Svrcek:2006yi,Arvanitaki:2009fg}. The de Broglie wavelength of the ultra-light scalar is so large that if it contributes to DM, it will prevent DM from clumping and forming too much structure on small scales~\cite{Hu:1998kj,Hwang:2009js} and, in this way, help with the $S_8$ problem.

In the $\Lambda$CDM model, there is only weak evidence that ultra-light scalars can make a significant difference when it comes to solving the $S_8$ problem~\cite{Hlozek:2017zzf,Lague:2021frh,Rogers:2023ezo}. On the other hand, already some years ago, it was pointed out by Allali, Hertzberg, and Rompineve \cite{Allali:2021azp} that adding a ULA to Cold NEDE leads to a simultaneous solution of the Hubble tension and the $S_8$ tension\footnote{A related phenomenological observation was made later in a different context in~\cite{Ye:2021iwa}.}; however, the ULA was not identified with the trigger-field, $\phi$, but was added as a new independent field. It is remarkable that Cold NEDE, already in its original form, holds in it this ingredient equivalent to a ULA, and all that is required is to relax the constraints on the initial $\phi$ value. No new fields are required.

In order to relax the constraints on the initial $\phi$ value, we needed to extend the Boltzmann code, which is being used to compute the CMB spectrum in the Cold NEDE model, to track the trigger field, $\phi$, and its perturbations after the phase transition and compute its effect on the final spectrum of CMB and density perturbations. With this paper, we therefore also provide an updated version of \textsc{TriggerCLASS}\footnote{\url{https://github.com/NEDE-Cosmo/TriggerCLASS}}, which is based on \textsc{CLASS} (the Cosmic Linear Anisotropy Solving System~\cite{Blas:2011rf}) and allows for sizeable trigger field oscillations.

The paper is split in a theoretical and data analysis part. In Sec.~\ref{sec:model}, we first review the Cold NEDE model at the background level emphasizing the role of the ultralight trigger. Then we outline a possible embedding of the model in an axion framework, which naturally explains the smallness of the trigger mass.  We next review the treatment of perturbations in the model and discuss the trigger sector, its dynamics and implementation in a Boltzmann code. We close the section by discussing the phenomenology of the Cold NEDE model. In section~\ref{sec:data}, we discuss the results of our Monte Carlo Markov Chain (MCMC) analysis by comparing Cold NEDE with $\Lambda$CDM for four different dataset combinations. This also includes a comparison with previous implementations of the model that assumed negligible trigger oscillations ($\Omega_\phi$ = 0). We conclude in Sec.~\ref{sec:conclusion}.

\section{Cold NEDE model}\label{sec:model}

The Cold NEDE model first proposed in \cite{Niedermann:2019olb} has been extensively described in \cite{Niedermann:2020dwg}. Here we highlight the most essential aspects and emphasize how it addresses the $S_8$ tension when the energy density of the trigger field is not neglected. In the Cold NEDE model, a first-order vacuum phase transition is triggered in the potential of the NEDE boson, $\psi$, by the dynamics of an ultralight trigger field, $\phi$, with mass $m$. Initially, the ultralight trigger field is frozen in its potential due to the Hubble friction, but as the Hubble rate decreases and the trigger becomes heavy compared to the Hubble scale, the trigger field starts to roll towards the minimum of its potential. When the trigger field rolls down, a new global minimum becomes accessible to the NEDE field $\psi$ via quantum tunneling, which triggers the phase transition. A general renormalizable potential of two scalar fields with the desired properties takes the form \cite{Niedermann:2019olb,Niedermann:2020dwg}
\begin{equation}
	V(\psi,\phi) = \frac{\lambda}{4}\psi^4 + \frac{1}{2}\be M^2\psi^2 - \frac{1}{3}\alpha M \psi^3 + \frac{1}{2}m^2\,\phi^2 + \frac{1}{2}\tilde{\lambda}\phi^2\psi^2 + \cdots\, ,
	\label{eq:coldNEDEpotential}
\end{equation}
where $\lambda$, $\tilde \lambda$, $\beta$ and $\alpha$ are dimensionless parameters, $M \sim \mathrm{eV}$ and the ellipsis represent possible higher order operators.
This potential is a general renormalizable two-field potential, but one can have other forms of the potential if one relaxes the condition that it should be renormalizable or allows the phase transition to be a second-order phase transition. However, since this is the form of the potential used in the original NEDE proposal, we will also restrict ourselves to this simple renormalizable form of the potential here.

\begin{figure}
\includegraphics[width=12cm]{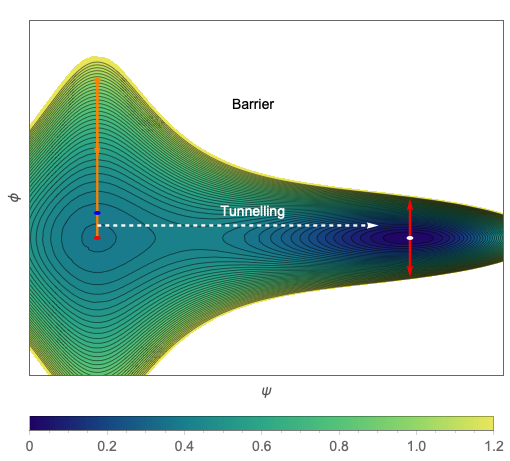}
\caption{Sketch of the two-field potential in \eqref{eq:coldNEDEpotential} (not to scale). The color coding corresponds to $[V(\psi,\phi)-V_\mathrm{min}]/[V(0,\phi_\mathrm{ini})-V_\mathrm{min}]$. The trigger field $\phi$ scans different profiles for the tunneling field $\psi$ as it rolls along the orange path. Tunneling becomes efficient between the blue and the red dot. The field finally settles in the true minimum (white dot). The $\phi$ direction corresponds to an ultralight field and contributes to DM at late times through its coherent oscillations (red arrow).  }
\label{fig:nedePotential3D}
\end{figure}

To have a first-order phase transition that is triggered as a second global minimum appears when $\phi$ begins to roll, we need to impose the conditions $\alpha^2 > 9 \be \lambda /2$ and $\be >0$. With these parameters, the evolution of the potential is as illustrated in Fig.~\ref{fig:nedePotential3D} \cite{Niedermann:2019olb, Niedermann:2020dwg}. When the trigger field, $\psi$, becomes heavy and rolls down along the orange path, it opens up the new vacuum for $\psi$. The NEDE boson $\psi$ then tunnels, and the system quickly undergoes a first-order phase transition.

The Euclidian action $S_E$ appearing in the exponent of the false vacuum decay rate $\Gamma = K \exp(-S_E)$ controls how quickly the phase transition happens where $K \sim M^4$ is the determinant factor.
As was demonstrated in \cite{Niedermann:2020dwg}, the Euclidian action can be written in the form\footnote{This result assumes a separation of scales where $\tilde \lambda / \lambda \ll 1$. As argued in \cite{Niedermann:2020dwg}, this reduces the 2D tunneling problem to a 1D tunneling problem in $\psi$.}
\beq\label{eq:SE}
S_{E} \approx \frac{4 \pi^2}{3\lambda} \left(2-\delta_\text{eff}\right)^{-3} \left(\alpha_1\delta_\text{eff}+\alpha_2\delta_\text{eff}^2+\alpha_3 \delta_\text{eff} ^3\right)~,
\eeq
where
\beq
\delta_\text{eff} = 9\frac{\lambda}{\alpha^2}\left(\beta+\tilde{\lambda} \frac{\phi^2(t)}{M^2}\right)~ ,
\eeq
and $\alpha_1 = 13.832$, $\alpha_2 = -10.819$, $\alpha_3 = 2.0765$ were determined numerically \cite{Adams:1993zs}.

For large initial values of the trigger field, $\phi$, \eqref{eq:SE} is not applicable, and tunneling is highly suppressed while $\delta_\text{eff} > 2$. As the Hubble rate drops and reaches the value\footnote{To be precise, in \cite{Niedermann:2020dwg}, it was found that $0.18 \lesssim H_* / m \lesssim 0.21$} $H = H_* \approx 0.2 ~m$, the Hubble friction becomes irrelevant, and the trigger field rolls quickly towards the minimum of its potential  $\phi\to 0$, driving $\delta_\text{eff}\to 9\lambda \be/\alpha^2 <  2$ and thus $S_E$ towards smaller values. For $\lambda < 0.1$, this makes the tunneling rate shoot up quickly and triggers an almost instantaneous phase transition at $H = H_*$. A  detailed analysis finds that the tunneling threshold is reached when $S_E \approx 250$ or  $\delta^*_\mathrm{eff} \approx 0.11 \times (\lambda / 0.01)$ (valid for $\lambda < 0.02$).

In previous work, we assumed that the energy density of the trigger field $\phi$ can always be neglected. At early times, that translates into the condition that $\phi_\text{ini}/M_\mathrm{Pl} \ll  1$, which can be satisfied simultaneously with the condition that makes sure that tunneling is initially blocked
\begin{align}\label{eq:bound_phi_ini}
\phi_\text{ini}^2 > \frac{2}{9} \frac{\alpha^2}{\lambda \tilde \lambda} M^2  \quad \left(\Rightarrow \delta_\mathrm{eff} > 2 \right) \,.%\gg \be M^2/\tilde\lambda
\end{align}
However, as discussed below, the condition that tunneling is initially blocked pushes us towards a regime of large initial $\phi$ values, where $\phi$ makes a small, yet non-negligible, contribution to the energy density. Therefore, in this work, we will relax the condition on the initial field value and just require $\phi_\text{ini}/M_\mathrm{Pl} \lesssim 1$. On a phenomenological level, this corresponds to introducing the (dimensionless) trigger relic density $\Omega_\phi = \rho_\phi|_\mathrm{today}/(3 M_\mathrm{Pl}^2 H_0^2$) as a new parameter. As we will show, this solves the $S_8$ problem and further improves the model's ability to solve the $H_0$ problem.

When the NEDE boson, $\psi$, is stuck in the false vacuum, it gives a new additional contribution to dark energy, which then decays away at the time of the phase transition $t_*$. The ratio of the NEDE contribution compared to the total energy density at the time of the phase transition is given by  $f_\text{NEDE} = \Delta V / \bar{\rho}(t_*)$ with $ \bar{\rho} $ being the total energy density and $\Delta V$ the released vacuum energy. From \eqref{eq:coldNEDEpotential}, we obtain
\begin{align}\label{def:f_NEDE}
f_\mathrm{NEDE} = \frac{c_\delta}{36} \frac{\alpha^4}{\lambda^3} \frac{M^4}{M_\mathrm{Pl}^2 H_*^2}\,,
\end{align}
where $c_\delta = (1+\sqrt{1-4\delta/9})^2 (1-2 \delta /3 + \sqrt{1-4\delta/9})/8$ and $\delta = 9 \beta \lambda / \alpha^2$. After tunneling, when the NEDE boson has picked up a non-vanishing vev, i.e.\ $\psi \to \psi_\mathrm{true}$, the last term in \eqref{eq:coldNEDEpotential} induces a shift in the trigger's mass,
\begin{align}\label{eq:shift_trigger_mass}
m^2 \to m^2 + \tilde \lambda \psi_\mathrm{true}^2\,.
\end{align}
Substituting this into \eqref{eq:bound_phi_ini} and using $\psi_\mathrm{True} = \frac{\alpha M}{2 \lambda} \left(1+\sqrt{1-4 \delta/9}\right)$ alongside \eqref{def:f_NEDE}, yields the lower bound
\begin{align} \label{eq:bound_phi_ini_2}
\frac{\phi_\text{ini}^2}{M_\mathrm{Pl}^2} >  \frac{H_*^2}{m^2} \frac{16}{\Delta m^2 / m^2} \frac{f_\mathrm{NEDE}}{(1-2 \delta /3 + \sqrt{1-4 \delta / 9})}
\end{align}

From previous work, we know that a resolution to the Hubble tension requires $f\nede\sim 0.15$, and the NEDE phase transition should happen just before matter radiation equality at a redshift $z_* \sim 5000$. We can relate this to the trigger mass through (assuming radiation domination)
\begin{align}\label{eq:m}
m = 1.7 \times 10^{-27} \, \text{eV} \, (1-f_\text{NEDE})^{-1/2}\left( \frac{g_*(T_*)}{3.4} \right)^{1/2} \left( \frac{z_*}{5000} \right)^2 \, \left(\frac{0.2}{H_*/m}\right) \,.
\end{align}
where $g_*$ is the effective number of relativistic degrees of freedom at the time of the decay.
The relation $H_* = 0.2 m$ then implies that $m$ is an ultralight mass scale of order $m \sim 10^{-27} \text{eV}$, while provided $0.01 \lesssim \lambda \lesssim 0.1$ and $\alpha \sim \beta \sim{\mathcal{O}(1)}$, the other mass scale, which sets the energy scale of the phase transition, is $M \sim  \text{eV}$ due to \eqref{def:f_NEDE}. As originally proposed, the hierarchy between the two mass scales can be explained in a UV completion with two axion fields, we will briefly elaborate below and return to in more detail in a future publication \cite{NEDE_axion}. Moreover, from \eqref{eq:bound_phi_ini_2}, we have $\Delta m^2/m^2 >  0.012 \times M_\mathrm{Pl}^2/\phi_\mathrm{ini}^2$, which shows that we should expect a negligible shift in the trigger mass for $\phi_\mathrm{ini}/ M_\mathrm{Pl} \gtrsim 0.1 $. We indeed find that the value of $\phi_\mathrm{ini}$ preferred by data is in interval $0.1 \lesssim \phi_\mathrm{ini}/M_\mathrm{Pl} \lesssim 1$.

At the time of the phase transition, $t_*$, the universe gets filled with small bubbles of true vacuum, which expand with the speed of light and start to collide on a time scale smaller than the Hubble time. Assuming that the phase transition is fast compared to the Hubble time (which can be easily achieved for $\lambda \lesssim 0.1$), the size of the bubbles as they collide corresponds to a physical size today, {$\ell_0 \ll \text{Mpc}$}, which is too small to be probed by the CMB. We will therefore ignore the small-scale anisotropy created by the bubbles and their collisions and treat the fluid of colliding bubbles as homogenous and isotropic fluid with adiabatic perturbations on large scales relevant for the CMB. The collision of bubbles is a complicated process, but on small scales, the colliding bubbles will carry anisotropic stress whose transverse and traceless part sources gravitational waves, while the remaining components source vector and scalar shear, which are known to behave like a stiff fluid\footnote{Another possibility for achieving an effective equation of state with $w_\mathrm{NEDE}> 1/3$ was discussed under ``scenario B'' in \cite{Niedermann:2021vgd}. The idea is to have the bubble wall condensate decay into a relativistic particle species that turns non-relativistic after a short time and contributes to dark matter. Here, the stiffer behavior arises due to a lack of dark matter at earlier times.} whose energy density, when averaged over large scales, decays as~\cite{Xue:2011nw} $1/a^6$. At the same time, the scalar condensate itself will also behave as a perfect fluid on large scales, whose equation of state will be ultimately determined by the EFT parameters in \eqref{eq:coldNEDEpotential}. Therefore, we expect the effective NEDE fluid after the phase transition to have a time-dependent equation of state $ w\nede(t) $ coming from different stiff and relativistic components. Here, instead of providing the full microscopic dictionary, we will treat $w_\mathrm{NEDE}$ as a phenomenological parameter.

We parametrize the background evolution of the initial false vacuum energy of the NEDE boson, $\psi$, decaying into an effective NEDE fluid of colliding bubbles with the form
\begin{equation}
	w\nede(t) = \begin{cases}
		-1 \hspace{1.35cm}\text{for}\quad  t < {t}_* \\
		w\nede(t) \quad \text{for}\quad t\geq {t}_*
	\end{cases},
\end{equation}
while the trigger field, $\phi$, evolves separately as an ultralight scalar field whose mass is fixed by the time of the phase transition, $z_*$, through $H_* = 0.2 m$. The other parameters relevant to the cosmological evolution are $w\nede$, $f\nede$, and $\Omega_\phi$. Provided we approximate $w\nede =\mathrm{const}$\footnote{This has to be viewed as the leading term in a Taylor expansion of $w_\mathrm{NEDE}(t)$ which determines the main effect on cosmological observables.}, the effective cosmological model has only four new d.o.f. compared to $\Lambda$CDM (the old EDE model \cite{Karwal:2016vyq} has three continuous and one discrete parameter), and all of them can be mapped one-to-one to the underlying theory parameters $\delta = 9\lambda\be/\al^2$ (the dimensionless parameter controlling the $\psi$ potential), $M$, $m$, and $\phi_\text{ini}$.

\subsection{Trigger as Ultra Light Axion}

Let us preview how the mass hierarchy between the $M$ scale and the $m$ scale can be understood in a UV completion using axions\footnote{This is closely related to the realization of hybrid inflation in two axion models recently studied in \cite{Kaloper:2020jso,Carta:2020oci}.}.
As we will show in a future publication, the model can be UV completed in a two-axion model, where the trigger field is an ultralight axion. In the limit where $\psi$ is small compared with its decay constant $\tilde f$, we can write the potential in the form (keeping only renormalizable terms in the $\psi$ expansion)
\beq
V(\psi,\phi) = \frac{1}{2}\lambda\psi^4 + \frac{1}{2}\be M^2 \psi^2 -\frac{1}{3}\alpha M \psi^3 -\Lambda^4\cos\left(\frac{\phi}{f}\right)-\frac{\tilde\Lambda^4}{ 2! \tilde f^2}\cos\left(\frac{\phi}{ f}\right)\psi^2+\frac{\tilde\Lambda^4}{4! \tilde f^4} \cos\left(\frac{\phi}{ f}\right)\psi^4 + \mathcal{O}(\psi^5)\,,
\label{eq:UVpotential}
\eeq
where as before $\lambda$, $\beta$ and $\alpha$ are dimensionless parameters, $M \sim \mathrm{eV}$ and $f$ is the axion decay constant associated with $\phi$. The parameters $\Lambda$ and $\tilde \Lambda$ carry mass dimension and can be argued to arise non-perturbatively from instanton corrections when a global shift symmetry is broken down to a discrete symmetry,  $\phi \to \phi + 2 \pi f$. In a nutshell, this makes $\Lambda, \, \tilde \Lambda \ll f,\, \tilde f$ a technically natural choice as we recover an enhanced symmetry in the limit $\tilde \Lambda, \Lambda \to 0$~\cite{tHooft:1979rat}.
For $\phi \ll f$, the above potential reproduces the usual Cold NEDE renormalizable potential in \eqref{eq:coldNEDEpotential}. The smallness of the trigger mass,  $m = \Lambda^2/f $, is then explained by the hierarchy $f \gg \Lambda$. From quantum gravity, we know that we must satisfy $f, \tilde f \lesssim M_p$. Taking $f \sim M_p$, then from $m = \Lambda^2/f = 10^{-27}$eV, we obtain $\Lambda \sim M \sim \mathrm{eV}$.

Similarly, from \eqref{eq:UVpotential}, we read off $\tilde \lambda =  \tilde\Lambda^4/(2 f^2\tilde f^2)$. At the same time, we need $\tilde \lambda \lesssim M^2/m^2 \sim 10^{-54}$ due to \eqref{eq:shift_trigger_mass} (assuming $\alpha, \lambda \sim 1$ for simplicity) in order to ensure a negligible shift in the trigger mass during the phase transition.\footnote{A similar bound, albeit somewhat weaker, arises from the requirement that radiative corrections to $m$, arising from loops of $\psi$, are suppressed. Using the same estimates as above, it was found in~\cite{Niedermann:2020dwg} that $\tilde \lambda \lesssim 10^{-50}$. This bound is relevant in the case where the trigger field contribution to dark matter can be neglected.} As before, the (extreme) smallness of $\tilde \lambda$ can be understood as a consequence of $\tilde f \gg \tilde \Lambda $; we satisfy the upper bound on $\tilde \lambda$ when $\tilde \Lambda \lesssim \sqrt{\tilde f M} $ is saturated. For completeness, we note that the parameters $\alpha$ and $\beta$ incur small shifts, which we can neglect.

To summarize, in this setup, the smallness of the trigger mass and its coupling to the NEDE boson can be understood from the trigger potential arising from the breaking of the global continuous shift symmetry, $\phi \to \phi + \mathrm{const}$, of an Axion-Like Particle (ALP) to a discrete shift symmetry, $\phi \to \phi + 2 \pi f$. The theory can then be UV completed to explain also the $\psi$ mass scale $M$ by promoting $\psi$ to an axion with discrete shift symmetry $\psi \to \psi + 2 \pi \tilde f$, as we will show in a future publication~\cite{NEDE_axion}.

\subsection{Cold NEDE field perturbations}

Before the NEDE phase transition, the false vacuum energy of the NEDE boson, $\psi$, behaves like a cosmological constant with negligible perturbations, but after the phase transition, the perturbations in the NEDE fluid can no longer be ignored. The trigger field, $\phi$, sets the initial condition for NEDE fluid perturbations after the phase transition. The adiabatic perturbations in the trigger field will make the phase transition happen at slightly different times in different places leading to small initial perturbations in the NEDE fluid after the phase transition. This picture is applicable on scales much larger than the typical bubble separation.

Treating the NEDE fluid as a perturbed perfect fluid, $\rho\nede (t, {\bf x}) = \bar\rho\nede (t) +\delta\rho\nede(t,{\bf x})$ and tracking the adiabatic perturbations, $\delta\phi$, on top of the background value of the trigger field, $\bar\phi$, the initial conditions for the NEDE fluid perturbations can be obtained using Israel junction conditions at the transition \cite{Niedermann:2020dwg}

\begin{subequations}
	\begin{align}
		\delta^*\nede &= -3 \left[1 + w\nede(\eta_*) \right] \mathcal{H}_* \frac{\delta \phi_*}{\bar{\phi}^\prime_*}\,,\\
		\theta\nede^* &= k^2 \frac{\delta \phi_*}{\bar{\phi}_*^\prime}\,,
	\end{align}
	\label{eq:matchingNEDE}
\end{subequations}
where $'$ denotes derivatives with respect to conformal time, $\eta = \int {\rm d}t/a$, and $\mathcal{H} = a H$. As usual, $\delta\nede \equiv \delta\rho\nede/\bar\rho\nede$ is the NEDE density contrast, $\theta\nede$ its velocity perturbation and the star denotes quantities which are evaluated at the time of the phase transition.  For a completely subdominant trigger field, i.e. $\Omega_\phi \ll \Omega_\mathrm{cdm}$, this is the only way the trigger leaves an imprint on cosmological observables. In particular, it can be shown that for adiabatic perturbations, the right-hand side of \eqref{eq:matchingNEDE} is independent of $\phi_\mathrm{ini}$.  The matching, therefore, gives rise to a rather unique fingerprint, which depends solely only on the dynamics of the (adiabatic) trigger.

Cold NEDE is a specific model in the NEDE framework.  The general NEDE framework describes a fast-triggered phase transition, where the trigger can be controlled by different dynamics in different model realizations. For example, in the Hot NEDE model, the trigger's role is instead given by the temperature of the dark sector\footnote{Therefore, the Hot NEDE model also does not contain an ultra-light trigger field. Instead, the Hot NEDE model contains interactions between dark matter and dark radiation, which can lead to a different simultaneous solution to the $H_0$ and $S_8$ tensions \cite{Niedermann:2021vgd,Niedermann:2021ijp} (see also \cite{Rubira:2022xhb}). }, corresponding to the replacement~\cite{Niedermann:2021vgd,Niedermann:2021ijp} $\phi \to T_d$ in \eqref{eq:matchingNEDE}. Since the trigger dynamics get imprinted in the perturbations of the NEDE fluid, future observations can discriminate between different NEDE models. How the different (N)EDE models respond to the $S_8$ tension is, in fact, already a first indication thereof.

The subsequent evolution of the NEDE fluid can then be computed with the usual equations governing the fluid perturbations \cite{Ma1995}, assuming vanishing viscosity and anisotropic stress on large scales relevant to the CMB. At the same time, it is assumed that the effective sound speed equals the adiabatic sound speed and that the equation of state after the decay is constant, so $c_s^2 = c_a^2 = w\nede$. We aim to check the robustness of these assumptions in our future work in two different ways. First by confronting the model with more constraining cosmological data and, second, by developing an explicit dictionary between the fluid description and the underlying microscopic model in \eqref{eq:coldNEDEpotential}. This treatment of the NEDE fluid was already implemented in previous versions of the Boltzmann code \textsc{TriggerCLASS}. However, below, we will introduce a new version of the code that also tracks the trigger field, $\phi$, \textit{after} the NEDE phase transition.

\subsection{Dark Matter Trigger dynamics and perturbations}

In previous studies of NEDE, it was assumed that the trigger field is completely subdominant. The only role of the trigger field, $\phi$, was to trigger the NEDE phase transition and set the initial conditions for the NEDE fluid perturbations after the transition (as we show below, this process is independent of $\phi_\mathrm{ini}$). For this reason, it was not necessary to track the evolution of the trigger field $\phi$ in the numerical Boltzmann code after the NEDE phase transition.
In this section, we discuss the trigger field in more detail and review a fluid approximation that allows us to track its perturbations numerically after the NEDE phase transition has taken place.

The trigger field, $\phi$, together with the potential in Eq.~\eqref{eq:coldNEDEpotential}, is described at the background level by
\begin{equation}
	\phi'' + 2{\cal H} \phi' + a^2\frac{{\rm d} V(\phi)}{{\rm d} \phi} = \phi'' + 2{\cal H} \phi' + a^2 m^2 \phi = 0,
	\label{eq:bkgTriggerEOM}
\end{equation}

This equation can be solved analytically in a radiation-dominated universe (approximately valid at the time of the phase transition),
\begin{align}\label{eq:phi_attr_parabolic}
\phi(x) = \sqrt{2} \, \Gamma(5/4) \, \phi_\text{ini} \, x^{-1/4} \,  J_{1/4}\left(x / 2\right) \,,
\end{align}
where $J_{1/4}$ is a Bessel function of the first kind, $x = m/H$ and $\Gamma$ is the Gamma function. The phase transition occurs as $\phi \to 0$ for the first time when $x^{-1} \approx 0.2$. Its energy density is $2 \rho_\phi = \phi'^2/a^2 + m^2 \phi^2$. For later times, i.e.\ $x \gg 1$, $\phi(x)$ becomes a quickly oscillating function and $\rho_\phi \propto 1/a^3$ when averaged over many cycles.  We can use this alongside \eqref{eq:bound_phi_ini_2} to estimate $\Omega_\phi$ today in terms of $\phi_\mathrm{ini}$ and $z_*$; explicitly (after setting $H_*/m=x_*^{-1}=0.2$, $g_* =3.4$ and using the estimates for $m$ given in \cite{Niedermann:2020dwg})
\begin{align}\label{eq:Omega_phi}
\Omega_\phi \approx 0.4 \times \left( \frac{1+z_*}{5000}\right) \left(\frac{\phi_\mathrm{ini}}{M_\mathrm{Pl}} \right)^2 \left( 1- f_\mathrm{NEDE}\right)\,.
\end{align}
We use this formula to guess the value of $\phi_\mathrm{ini}$ for a given target value of $\Omega_\phi$, as required by the ``shooting algorithm'' we implemented in \textsc{TriggerCLASS}. Thereafter, the actual numerical integration solves \eqref{eq:bkgTriggerEOM} directly. As a reference point, we used $\phi_\mathrm{ini} = 0.0001 \times M_\mathrm{Pl}$ in our previous {\tt CLASS} implementations of Cold NEDE, which indeed leads to a strong suppression $\Omega_\phi \ll 1$.

The evolution equation for the trigger field perturbations in synchronous gauge is
\begin{equation}
	\delta\phi'' = - 2 {\cal H} \delta\phi' - \left(k^2 + a^2 \frac{{\rm d}^2 V}{{\rm d}\phi^2} \right) \delta\phi - \frac{h'}{2}\phi',
	\label{eq:pertsTrigger}
\end{equation}
where $h$ is the spatial trace of the metric perturbation. The trigger perturbations are sourced by curvature perturbations $\propto h$ (to be specific, on super-horizon scales $h = -\frac{\zeta}{2} (k \eta)^2 + \mathcal{O}(k^4 \eta^4)$, where $\zeta$ is the curvature mode).  In the adiabatic case, the homogeneous part of the solution is set to zero such that it is solely fixed in terms of the curvature perturbation. As the sourcing term is controlled by the background solution in \eqref{eq:phi_attr_parabolic}, $\delta \phi \propto \phi_\mathrm{ini}$, and thus $\phi_\mathrm{ini}$ indeed drops out of the matching in \eqref{eq:matchingNEDE}. Its main observable effect is, therefore, to dial the amount of trigger DM through \eqref{eq:Omega_phi}.

Computing the background field and perturbations using the equations above during the whole evolution is a numerically expensive task. Generally, it requires an increased precision because of fast oscillations in the case of ULA's with masses that are well below an inverse Hubble time today of $H_0\simeq 10^{-33}$ eV. This situation requires to switch to a fluid approximation at an intermediate point of the computation when the Hubble friction term is small compared to the mass of the trigger field, i.e., $H/m \ll 1$ \cite{Hu:1998kj, Poulin:2018dzj}. An effective fluid description is necessary but can, in its simplest version, lead to errors of up to 2$\sigma$ \cite{Cookmeyer:2019rna} for a trigger field with a mass around $m \sim 10^{-27}$. Often in the literature dealing with ULA's, terms of order $H/m$ are ignored in the effective sound speed in the fluid approximation and the matching of the field evolution to the fluid approximation. This omission is the source of the relatively large errors. The implementation used here, identical to the one suggested in \cite{Passaglia:2022bcr}, is higher-order in $H/m$ leading to a fluid approximation with negligible errors. In a nutshell, it consists in introducing the fluid variables $\rho^\mathrm{efa}_\phi$ and $p^\mathrm{efa}_\phi$, which are matched to the cycle-averaged trigger energy density and pressure at some time $t_m > t_*$. The superscript ``${\rm efa}$'' stands for \emph{effective fluid approximation}, and further details can be found in Appendix~\ref{app:matching}.

At background level, the initial value $\rho^\mathrm{efa}_\phi|_m$  is then evolved  using the continuity equation,
\begin{equation}
	 \frac{d \rho^{\rm efa}_\phi}{d \eta} = - 3 \mathcal{H} \left(\rho^{\rm efa}_\phi + p^{\rm efa}_\phi\right).
	\label{eq:efContinuity}
\end{equation}

As usual, the system is closed by specifying the equation of state parameter $p^{\rm efa}_\phi/\rho^{\rm efa}_\phi \equiv w^{\rm efa}_\phi$. We follow the suggestion in  \cite{Passaglia:2022bcr} and use  $w^{\rm efa}_\phi = (3/2)(H/m)^2$. This coincides with what can be obtained from a WKB approximation at second order from the trigger's equation of state during radiation domination. This is an improvement over the approximations used in the past, which led to a larger evolution error.

For the perturbations, there is an interplay of two scales, the first is given by $m$, already present at the background level, and the second one is given by the momentum $k$ of a given mode. Applying a WKB approximation to Eq.~\eqref{eq:pertsTrigger} and ignoring the metric sourcing, one obtains
\begin{equation}\label{eq:k_J}
	\delta\phi \propto \cos\left[m \int a \mathrm{d}\eta  + k \int c_{s,\phi} {\rm d}\eta + \varphi_0\right],
\end{equation}
where $\varphi_0$ is a phase, and the sound speed is given by
\begin{equation}
	c_{s,\phi} = \left(\frac{k}{am}\right)^{-1}\left( \sqrt{1+\left( \frac{k}{am} \right)^2} - 1\right) .
\end{equation}

The dependence of the effective sound speed on the wave number, $k$, is crucial to suppress the linear matter power spectrum at small scales below the Jeans length $2 \pi/k_\mathrm{J, eq} \sim 1/\sqrt{m \, a_\mathrm{eq}^2 H_\mathrm{eq}}$ \cite{Hu:2000ke, Arvanitaki:2009fg,Hwang:2009js}. In fact, it has been estimated that the scales probed by the $S_8$ parameter can be suppressed by around $5-10\%$ \cite{Kobayashi:2017jcf, Allali:2021azp}. On the other hand, for larger scales than the Jeans length (when the second term in \eqref{eq:k_J} is negligible), such an ultra-light field behaves as cold dark matter. In our case with a trigger mass of around $10^{-27}$ eV, the suppression kicks in at  comoving wave numbers larger than
\begin{equation}
	k_{\rm J, eq} \simeq 0.16 \; {\rm Mpc}^{-1} \left(\frac{m}{10^{-27}\; {\rm eV}\;}\right)^{1/2}.
\end{equation}
These estimates indeed agree with our finding in the right panel of Fig.~\ref{fig:Pk_NEDE}, where we compare the linear power spectrum in NEDE for $\Omega_\phi= 0 $ and  $\Omega_\phi \neq 0 $ for our best-fit cosmologies.

We can again use a fluid approximation to avoid the quick oscillations arising from the first term in \eqref{eq:k_J}. Crucially, the Jeans oscillations corresponding to the second term in \eqref{eq:k_J} are still captured by the fluid description. To that end, we introduce the density fluctuations, $\delta^{\rm efa}_\phi = \delta\rho^{\rm efa}_\phi/ \rho^{\rm efa}_\phi$ and the velocity divergence, which as before are matched at time $t_m$ (see Appendix~\ref{app:matching}).  They are evolved according to~\cite{Hu:1998kj}
\begin{subequations}
	\begin{align}
	\delta'^{\rm efa}_\phi &= -(1 + w^{\rm efa}_\phi)\left( \theta^{\rm efa}_\phi + \frac{h'}{2}\right) - 3(c_s^2 - w^{\rm efa}_\phi){\cal H} \delta^{\rm efa}_\phi - 9(1+ w^{\rm efa}_\phi)(c_s^2 - c_a^2){\cal H}^2 \frac{\theta^{\rm efa}_\phi}{k^2},\\
	\theta'^{\rm efa}_\phi &= -(1-3c_s^2) {\cal H}\theta^{\rm efa}_\phi + \frac{c_s^2 k^2}{1 + w^{\rm efa}_\phi}\delta^{\rm efa}_\phi
	\end{align}
\end{subequations}
where $c_a$ is the adiabatic sound speed,
\begin{equation}
	c_a^2  = w^{\rm efa}_\phi - \frac{w'^{\rm efa}_\phi}{3(1+w^{\rm efa}_\phi){\cal H}}.
\end{equation}
Moreover we follow the study in~\cite{Passaglia:2022bcr} and define the effective sound speed as
\begin{equation}
	c_s^2 = c_{s,\phi}^2 + \frac{5}{4}\frac{H^2}{m^2},
\end{equation}
where the second term has been determined numerically to reduce the evolution error below $\mathcal{O}(H^2/m^2)$.

\subsection{NEDE phenomenology}

\begin{figure}
     \centering
          \begin{subfigure}[b]{0.49\textwidth}
         \centering
         \includegraphics[width=\textwidth]{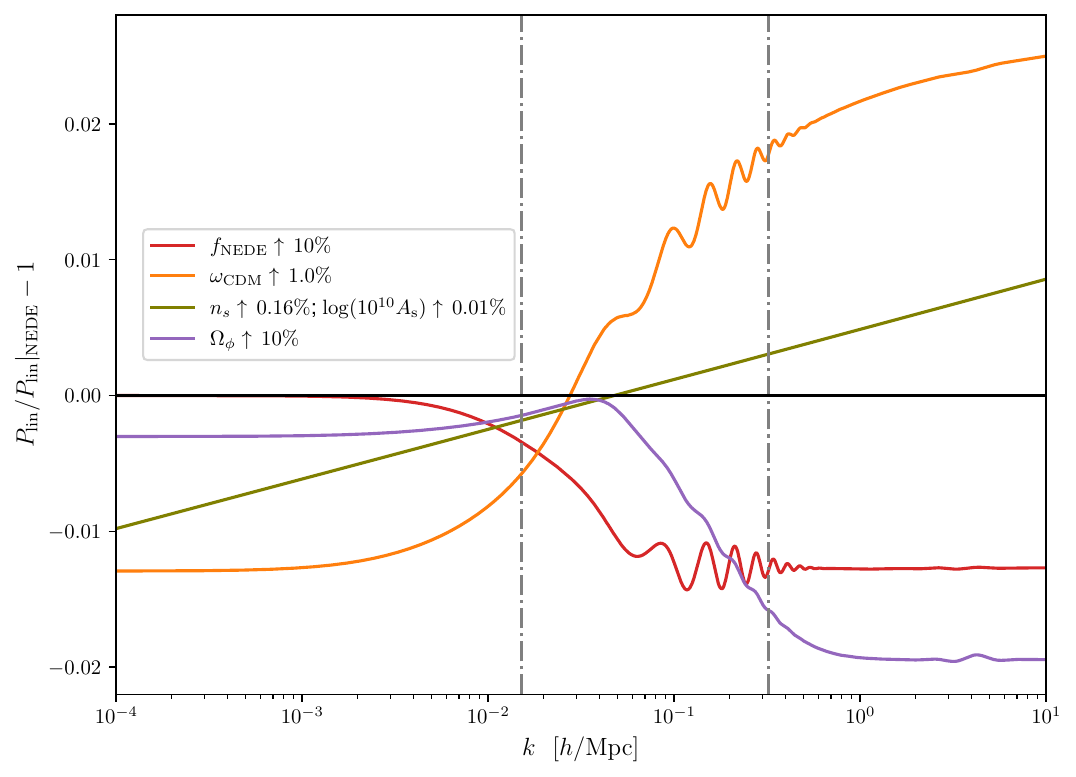}
         \caption{Sensitivity to NEDE parameters }
         \label{fig:Pk_vary_params}
     \end{subfigure}
          \hfill
     \begin{subfigure}[b]{0.49\textwidth}
         \centering
         \includegraphics[width=\textwidth]{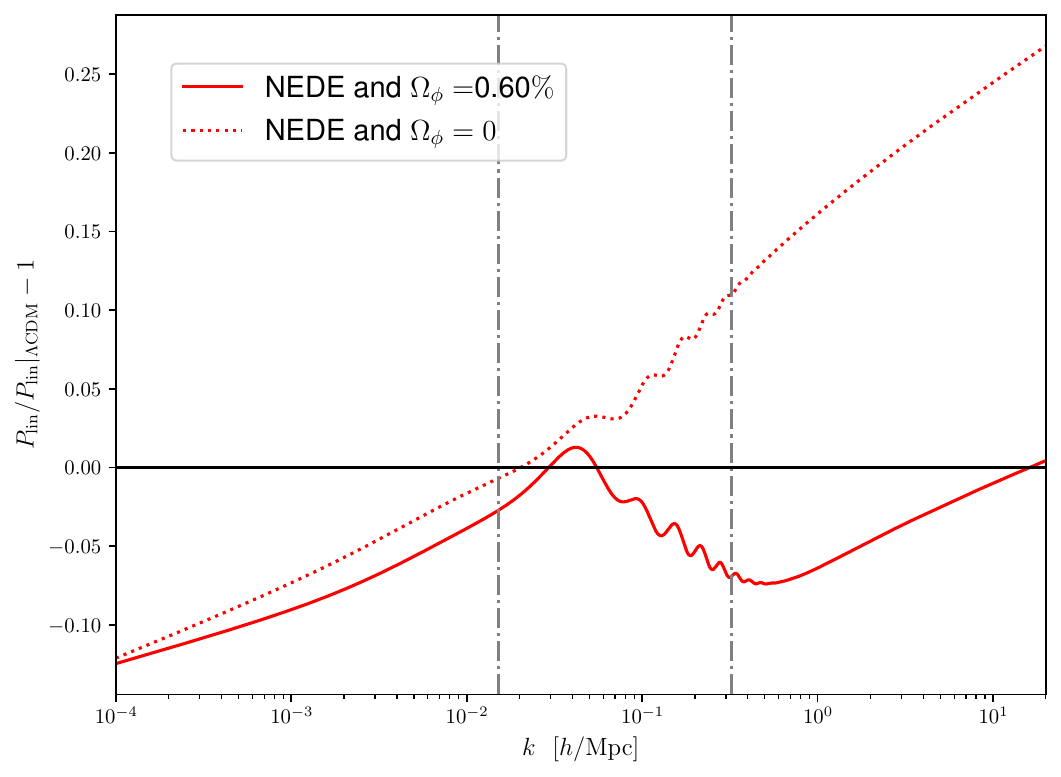}
         \caption{ $\Lambda$CDM versus NEDE}
         \label{fig:Pk_compare}
     \end{subfigure}
        \caption{Analysis of the linear matter power spectrum $P_\mathrm{lin}$ as inferred from the baseline + $S_8$ + $H_0$ dataset combination. The left panel (a) shows how the NEDE prediction changes as a single parameter ($f_\mathrm{NEDE}$, $w_\mathrm{cdm}$, $\Omega_\phi$, $n_s$ and $A_s$) is varied relative to the best-fit in Tab.~\ref{tab:paramResults}. The changes are chosen to reflect the approximate degeneracies in the parameters. The right panel (b) depicts the relative difference between $\Lambda$CDM and NEDE with $\Omega_\phi = 0$ (red, dotted) and $\Omega_\phi \neq 0$ (red, solid). The dash-dotted lines delineate the range where the $\sigma_8$ integration in \eqref{sigma8} picks up $90 \%$ of its value.}
        \label{fig:Pk_NEDE}
\end{figure}

The phenomenological success of NEDE relies on the presence of different approximate degeneracies between $f_\mathrm{NEDE}$ and the parameters $H_0$, $\omega_\mathrm{cdm}$ and $n_s$. For example, the main degeneracy shared by all EDE type models is between $f_\mathrm{NEDE}$ and $H_0$. In short, increasing $f_\mathrm{NEDE}$ reduces characteristic scales in the primordial fluid, such as the sound horizon, which on the level of the observed angular scales can be compensated by increasing $H_0$. Overall, these degeneracies ensure that the CMB fit can be preserved (and even improved) despite the fact that NEDE introduces an order $10\%$ change to the energy budget during the CMB epoch when we would have naively thought the budget to be constrained at least at the percent level.  These effects have been discussed in detail in \cite{Niedermann:2020dwg} in terms of the gravitational potential and the TT power spectra.  As we are mostly concerned with the $S_8$ tension in this work, we will instead focus on the effect NEDE has on the linear matter power spectrum $P_\mathrm{lin}(k)$. To be specific, $S_8 = \sigma_8 \sqrt{\Omega_m/0.3}$ depends on $P_\mathrm{lin}$ through $\sigma_8$, which is the root-mean-square mass fluctuation in a sphere of comoving radius $8 \mathrm{Mpc}/h$,
\begin{align}\label{sigma8}
\sigma_8^2 = \frac{1}{2 \pi^2} \int \mathrm{d} k  k^2 P_\mathrm{lin}(k) W^2(k \times 8 \mathrm{Mpc}/h),
\end{align}
where we introduced the Fourier transform of a top-hat window function, $W(x) = (3/x) [ \sin(x) / x^2 - \cos(x) /x ] $.
As we will see, the changes that nearly preserve the fit to the CMB are still leaving an imprint on the matter power spectrum, making it an important probe to test the model.
To better understand how individual parameters affect the matter power spectrum, we show in Fig.~\ref{fig:Pk_vary_params} how $P_\mathrm{lin}$ responds to changes in $f_\mathrm{NEDE}$, $\omega_\mathrm{cdm}$, ($n_s$, $A_s$), and $\Omega_\phi$ relative to the NEDE best-fit cosmology. The parameter variations are chosen such that they reflect the typical NEDE degeneracies (see also Fig.~\ref{fig:nedeTriangle}). Increasing $f_\mathrm{NEDE}$ leads to an excess decay of the gravitational potential~\cite{Lin:2019qug, Vagnozzi2021,Niedermann:2020dwg}. Due to the reduced gravitational attraction, this implies a power deficit in $P_\mathrm{lin}$ on small scales (see the red curve for $k > 0.1 h / \mathrm{Mpc}$). This can be approximately (yet not entirely) compensated by increasing the amount of dark matter $\omega_\mathrm{cdm}$ (yellow curve) as well as $n_s$ and  $A_s$ (blue curve). On large scales ($k \ll 0.1 h / \mathrm{Mpc}$), increasing $w_\mathrm{cdm}$ leads to a power deficit, which can be explained through changes in the background evolution (matter modes enter the horizon later and as a result, have less time to grow). This effect remains largely uncompensated.

The net effect is then shown in Fig.~\ref{fig:Pk_compare} where we compare NEDE with a $\Lambda$CDM cosmology (constrained with the same data sets). With $\Omega_\phi = 0$, there is a power deficit on large scales and an excess on small scales. As the value of $S_8$ is mostly sensitive to the excess (the integral in \eqref{sigma8} picks up its main contribution between the dashed lines), there is an increase of $S_8$ compared to $\Lambda$CDM. This observation has been used in the past to argue against EDE-type solutions to the Hubble tension~\cite{Hill:2020osr, Ivanov:2020ril, DAmico:2019fhj}. However, things change dramatically for $\Omega_\phi > 0 $. Now the Jeans (or acoustic) oscillations in the trigger dark matter component damp the power spectrum on exactly the scales relevant for $S_8$, leading to a net suppression. This explains why NEDE can help with the $S_8$ tension. We further note that despite $\Omega_\phi > 0$, the power spectrum is still enhanced on sub-$S_8$ scales with $k > 10 h /\mathrm{Mpc}$. As pointed out in the context of EDE [80], this affects galaxy formation processes, and we speculate that the bluer spectrum on small scales might help explain recent data from the James Webb Space Telescope that suggests the presence of high-redshift galaxies with unexpectedly high stellar masses~\cite{Boylan-Kolchin:2022kae, Parashari:2023cui} (see however ~\cite{McCaffrey:2023qem} for an opposing claim).

\section{Data analysis}\label{sec:data}

To understand the impact of tracking the trigger evolution, we perform several Markov Chain Monte Carlo (MCMC) sampling runs, comparing the $\Lambda$CDM model, the NEDE model where the trigger field is not tracked after the phase transition and thus does not act as dark matter ($\Omega_\phi = 0$) and the NEDE including the trigger evolution ($\Omega_\phi>0$). We take the following datasets to be our baseline:
\begin{itemize}
	\item \textbf{Planck 2018}~\cite{Aghanim2020} We use the complete CMB likelihoods reported by the Planck collaboration in 2018. This includes the TT, TE, EE, as well as the lensing power spectrum, containing information for multipoles between $\ell_{\rm min} = 2$ and $\ell_{\rm max}=2000$. ($\ell_{\rm max}=2500$ for the TT spectrum only).

	\item \textbf{BAO}: We include the baryon acoustic oscillations (BAO) measurements performed at different redshifts. Specifically, we include the results of the Six-Degree-Field Galactic Survey (6dFGS)~\cite{Beutler2011} with a median redshift of $z=0.052$, the clustering data from the Main Galactic Survey (MGS) coming from the Sloan Digital Sky Survey (SDSS)\cite{Ross2015} with information on $z\lesssim 0.2$ and the BAO-only likelihood from the Baryon Oscillation Spectroscopic Survey (BOSS), which is part of the SDSS-III~\cite{Alam2017}, and includes luminous red galaxies up to $z\approx 0.4$ (LOWZ) and massive galaxies between $0.4<z<0.7$ (CMASS).

	\item \textbf{Pantheon SN Ia Samples}: Likelihood from the sample of 1048 spectroscopically confirmed Type Ia supernovae~\cite{Scolnic2018} recorded at redshifts between $0.01< z < 2.26$ and their corresponding distance modulus. When including S$H_0$ES, we include a prior for $H_0$ and do not sample the absolute magnitude parameter.

\end{itemize}

In addition to our baseline, we do an analysis where we include a large-scale structure (LSS) constraint and/or the local $H_0$ measurement constraint from 2021. Since these are the datasets with which $\Lambda$CDM is in tension, we study the NEDE model in its previous ($\Omega_\phi = 0$) and current implementation ($\Omega_\phi\neq 0$) focusing on these datasets.

\begin{itemize}
	\item \textbf{LSS}: For the MCMC chains, we impose the LSS constraint from \cite{Joudaki2020} resulting from the combined analysis of weak gravitational lensing tomographies from the Kilo Degree Survey (KiDS) + VIKING (HV450)  and the Dark Energy Survey (DES) Y1:
	\begin{equation}
		S_8 = 0.762^{+0.025}_{-0.024},
		\label{eq:S_8combinedConst}
	\end{equation}
	although newer non-combined constraints exist that mildly modify the central value and barely reduce the error bars (see \cite{Asgari2021, DES:2021wwk}).
	\item \textbf{SH$0$ES}: We include a prior directly on $H_0$, specifically we employ
	\begin{equation}
		H_0 = 73.04\pm 1.04 \;{\rm km/s/Mpc},
	\end{equation}
	from the latest report by the SH$0$ES collaboration~\cite{Riess2022}, which reaches into the Hubble flow by means of a distance ladder based on Cepheid stars and Supernovae Ia.
\end{itemize}

The MCMC samplings were run until they reached a Gelman-Rubin criterion value of $R-1< 0.03$. They were performed by using \textsc{TriggerCLASSv6.1} together with the Bayesian sampler \texttt{Cobaya}\cite{Torrado2021}\footnote{\url{https://cobaya.readthedocs.io/en/latest/index.html}}. All chains were produced with standard settings, namely uniform priors on sampled parameters, two massless neutrinos and one massive neutrino with $m_{\rm ncdm} = 0.06\, {\rm eV}$, default temperature of $T_{\rm ncdm} = 0.71611 \; T_{\rm cmb}$. Moreover, the viscosity parameter of the NEDE fluid was set to zero, and the rest-frame sound speed was assumed to equal the adiabatic sound speed ($w\nede = c_s^2 = \mathrm{const}$). We further fixed $H_*/m = 0.2$ following the derivations in \cite{Niedermann:2020dwg}. We sample the original six parameters already present in the $\Lambda$CDM model, $(\omega_b, \omega_{\rm c}, H_0, \tau_{\rm reio}, \log(10 A_{\rm s}), n_s)$, the additional parameters $(f\nede, w\nede,\log_{10}(z_*))$ in the basic NEDE simulations with $\Omega_\phi = 0$ and include {$\Omega_\phi$} when considering the trigger as DM. In the latter case, we also include $\phi_\mathrm{ini}$ (alongside $m$) as a derived parameter.

The main purpose of the simulations performed is to verify the suppression of power in the linear matter spectrum for large wave numbers/small scales coming from the ultra-light trigger field perturbations described in the previous section, thus reducing the $S_8$ tension. Across the board, we observe that the inclusion of weak lensing data (LSS) favors that a non-zero fraction of the energy budget today belongs to the trigger field, $\phi$, corresponding to a weak $1-2 \sigma$ evidence and a posterior mean of about $\Omega_{\phi} = 0.5\%$. Specifically as can be seen in Table~\ref{tab:paramResults}, for the run combining the baseline datasets and LSS prior, we obtain $\Omega_{\phi} = 0.45^{+0.17}_{-0.37}\;\%$ and for the MCMC chains produced with the baseline, S$H_0$ES and LSS we get $\Omega_{\phi} = 0.57^{+0.25}_{-0.29}\; \%$. Moreover, in all runs where the trigger field is tracked and contributes to dark matter, one obtains posteriors, and best fits for $S_8$, which are consistently lower compared to the analogous MCMC runs where the trigger field is neglected after the NEDE transition. Specifically, in the NEDE runs, the means for $S_8$ decrease by approximately $1\sigma$ for the baseline run and by close to $3 \sigma$ for the run with all datasets included.

\begin{figure}[t]
	\centering
	\includegraphics[clip,width=\textwidth]{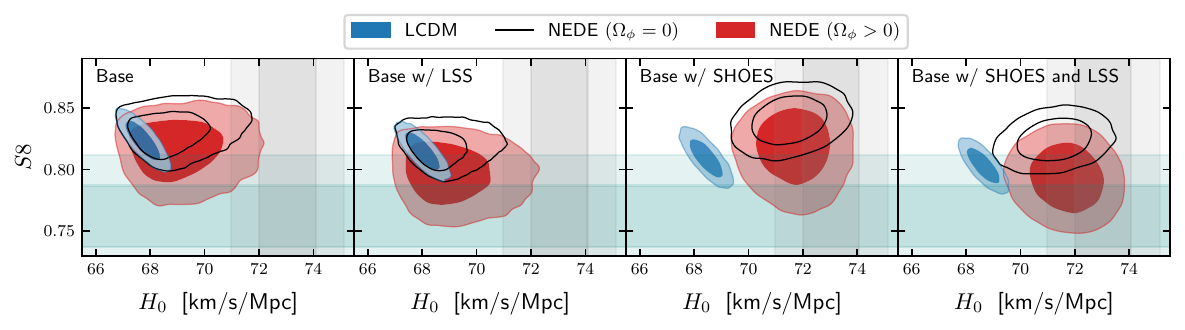}
	\caption{Comparison of $H_0$ vs $S_8$ contours for the $\Lambda$CDM model and Cold NEDE with $\Omega_\phi =0$ (black) and $\Omega_\phi >0$ (red). The horizontal and vertical bands correspond to the $1\sigma$ and $2\sigma$ standard deviations of the constraint on $H_0$ from the S$H_0$ES collaboration~\cite{Riess2022} and the one on $S_8$ from the combined analysis of KiDS+Viking+DES~\cite{Joudaki2020}. Here and henceforth dark and light shaded regions correspond to the $68 \%$ and $95 \%$ confidence limit (C.L.), respectively.}
	\label{fig:S_8H0plot}
\end{figure}

The highlight of the NEDE model is its ability to simultaneously reduce the Hubble tension and the $S_8$ tension. We can witness this in different ways. Let us first discuss the differences in the $S_8-H_0$ plane for the three models considered: $\Lambda$CDM, NEDE $(\Omega_{\phi}=0)$ and NEDE $(\Omega_{\phi}>0)$. The results of the MCMC chains are shown in Fig.~\ref{fig:S_8H0plot}. To begin with, we can see that the $\Lambda$CDM contours are somewhat unaffected by the inclusion of LSS or S$H_0$ES and remain relatively far from the intersection of the LSS and S$H_0$ES constraints, reflecting the existing tensions. NEDE $(\Omega_{\phi}=0)$ on the other hand, as has been concluded previously \cite{Niedermann:2020dwg, Cruz:2022oqk}, reduces the tension with S$H_0$ES to about $2\sigma$ while enhancing the power of the linear power spectrum at the scales relevant for $\sigma_8$, mildly worsening the $S_8$ tension. This is reflected by the black contours, which for all data set combinations remain above the LSS 1$\sigma$ band and barely overlap the $2\sigma$ band for some data set combinations. Remarkably, the NEDE model, when including the full evolution of the trigger ($\Omega_\phi > 0$), produces a contour which overlaps with the $1-2\sigma$ bands of both LSS and S$H_0$ES using only the baseline datasets. Importantly, this justifies the combined analyses, including Gaussian priors on LSS and $H_0$ shown in the panels to the right. Interestingly, adding only the $H_0$ prior (third panel) leads to a slightly reduced $S_8$ tension. In other words, within NEDE, both datasets ``pull'' in the same direction provided $\Omega_\phi$ is allowed to vary freely.

The results described can be quantitatively captured by the improvement in the total $\chi^2$ values at the best-fit set of parameters. In Table~\ref{tab:chiSqDiffNEDEOmega0vsLCDM} and Table~\ref{tab:chiSqDiffNEDEvsLCDM}, we show the individual $\chi^2$ differences with respect to $\Lambda$CDM, as well as the total differences and the remaining tensions, for NEDE $(\Omega_{\phi}=0)$ and NEDE $(\Omega_{\phi}>0)$, respectively. In Table~\ref{tab:paramResults}  we state the best-fit, posterior mean, and standard deviation of the relevant parameters for $\Lambda$CDM and the two variations of NEDE considered for this study.

Firstly, Table~\ref{tab:chiSqDiffNEDEOmega0vsLCDM} covers the case with $\Omega_\phi=0$ and shows how we recover known improvements in $\chi^2$ compared to $\Lambda$CDM, as in \cite{Niedermann:2020dwg, Cruz:2022oqk}. The best improvement corresponds to the data set combination baseline + S$H_0$ES. In this particular case, the model reduces the Hubble tension very significantly to $1.32\sigma$. This uses the \textit{difference of the maximum a posteriori} (DMAP) tension measure~\cite{Raveri:2018wln}, given by
\begin{align}
Q_{\rm DMAP}(H_0) &= \sqrt{\chi^2_{\rm M + SH0ES} - \chi^2_{\rm M}},\\[5pt]
Q_{\rm DMAP}(S_8) &= \sqrt{\chi^2_{\rm M + LSS} - \chi^2_{\rm M}},
\end{align}
for the $H_0$ and $S_8$ tensions respectively and where ${\rm M}$ is a data set combination not involving the tension. This measure has been argued to be best adapted to assessing EDE type models~\cite{Schoeneberg2022}. In fact, this is even better than our previously reported result, which can be explained with the baseline run giving a better total $\chi^2$ value here, as the equation of state was not sampled in~\cite{Niedermann:2020dwg, Cruz:2022oqk}. We also note that including the LSS likelihood in the MCMC runs for NEDE $(\Omega_\phi = 0)$ leads to a smaller overall $\chi^2$ improvement, reflecting some degree of mutual incompatibility between resolving the Hubble and the $S_8$ tension in the reduced model. In particular, the $S_8$ tension is increased in both the $\Lambda$CDM and the NEDE $(\Omega_\phi = 0)$ model when the S$H_0$ES prior is included.

In addition to the DMAP tension measure, the last two rows of Table~\ref{tab:chiSqDiffNEDEOmega0vsLCDM} contain the Gaussian tension computed as
\begin{equation}
	%x_{\rm Gaussian Tension} =
	 \frac{|(\bar{H}_0)_\mathrm{base} - (\bar{H}_0)_{\mathrm{SH}_0\mathrm{ES}}|}{\sqrt{(\sigma^2_{H_0})_{\mathrm{base}} + (\sigma^2_{H_0})_{\mathrm{SH}_0\mathrm{ES}}}}
	 \quad \text{and} \quad
	 \frac{|(\bar{S}_8)_\mathrm{base} - (\bar{S}_8)_{\mathrm{LSS}}|}{\sqrt{(\sigma^2_{S_8})_{\mathrm{base}} + (\sigma^2_{S_8})_{\mathrm{LSS}}}}
\end{equation}
for $H_0$ and $S_8$, respectively. Here, we denote the mean of a quantity $x$ as $\bar{x}$ and its variance as $\sigma_x$. Applying this measure directly to our runs, yields a still significant $H_0$ tension at the $3 \sigma$ level. This might appear surprising given the much smaller tension when considering the DMAP measure. However, we can attribute it to well-known sampling volume effects~\cite{Niedermann:2020dwg}. In short, in the limit where $f_\mathrm{NEDE} \to 0$, the parameters $\log_{10}(z_*)$ and $ w_\mathrm{NEDE}$ become unconstrained. This enhances the sampling volume and leads to non-Gaussian artifacts in the posterior for $f_\mathrm{NEDE}$ and $H_0$. In particular, it favors smaller values of $H_0$. This can be addressed by fixing the trigger mass $m$ (or $z_*$ alternatively) alongside $w_\mathrm{NEDE}$, which has been demonstrated to give compatible results with both the DMAP tension and an independent profile likelihood approach~\cite{Cruz:2023cxy} (for the case of EDE see~\cite{Herold:2021ksg}). To demonstrate this further, we provide a full triangle plot of an example run with fixed $z_*$ and $w\nede$ in Appendix \ref{app:triangles}. While the $\Omega_\phi=0$ model thus provides a solution to the Hubble tension, it cannot address the $S_8$ tension. In fact, we observe that the latter remains between $2-3\sigma$ and is even increased when the fraction of the NEDE fluid increases (by including an $H_0$ prior).

\begin{table}[htbp]
	\centering
	\begin{tabular}{>{\arraybackslash}p{4cm}|>{\centering\arraybackslash}p{2.5cm}|>{\centering\arraybackslash}p{2.5cm}|>{\centering\arraybackslash}p{2.5cm}|>{\centering\arraybackslash}p{2.5cm}}
		\centering NEDE $(\Omega_\phi = 0)$   & Baseline     & + LSS        & + $H_0$      & + $H_0$ + LSS \\ \hline\hline
		Pl.2018 low $\ell$ TT                 & -1.4         & -0.2         & -1.5         & -1.7          \\ \hline
		Pl.2018 low $\ell$ EE                 & 0.4          & -1.5         & -0.9         & 0.6           \\ \hline
		Pl.2018 high $\ell$ TTTEEE            & -1.3         & -0.2         & -5.2         & -2.7          \\ \hline
		Pl.2018 Lensing                       & 0.3          & 0.4          & 0.5          & 0.5           \\ \hline
		BAO SDSS DR7 mgs                      & 0.6          & -0.3         & -0.5         & 0.1           \\ \hline
		BAO 6dF 2011                          & -0.1         & 0.0          & 0.0          & 0.0           \\ \hline
		BAO SDSS DR12                         & -1.5         & 0.3          & 0.0          & 0.2           \\ \hline
		SN Pantheon                           & -0.4         & 0.1          & 0.1          & 0.1           \\ \hline
		S$H_0$ES 2021                         & --           & --           & -16.4        & -16.8         \\ \hline
		LSS                                   & --           & 0.3          & --           & 4.6           \\ \hline\hline
		\centering $\Delta\, \chi^2$          & -3.4         & -1.1         & -23.9        & -15.1         \\ \hline
		\centering $H_0-Q_{\rm dmap}$ Tension & --           & --           & 1.32         & 1.97          \\ \hline
		\centering $S_8-Q_{\rm dmap}$ Tension  & --           & 2.74         & --           & 3.11          \\ \hline
		\centering $H_0$ Gauss.Tension        & 3.53$\sigma$ & 3.79$\sigma$ & --           & --            \\ \hline
		\centering $S_8$ Gauss.Tension         & 2.45$\sigma$ & --           & 2.77$\sigma$ & --            \\ \hline
	\end{tabular}
	\caption{Table of $\Delta\chi^2= \chi^2_{_{{\rm NEDE} (\Omega_{\phi}=0)}}\!\! - \chi^2_{\Lambda{\rm CDM}}$ differences between the NEDE $(\Omega_\phi = 0)$ and $\Lambda$CDM aggregated by datasets.}
	\label{tab:chiSqDiffNEDEOmega0vsLCDM}
\end{table}

Next, Table~\ref{tab:chiSqDiffNEDEvsLCDM} shows the $\chi^2$ differences of the NEDE ($\Omega_\phi \neq 0$) model when sampling its four parameters $(f\nede, \log_{10}(z_*), w\nede, \Omega_{\phi})$. The columns, including the LSS likelihood, show gains (i.e.\ $\chi^2$ decrements) in almost all individual likelihoods. Moreover, the total $\chi^2$ improves over the NEDE $(\Omega_{\phi}=0)$ version described before for all dataset combinations (compare Table~\ref{tab:chiSq-NEDEomega0} and \ref{tab:chiSq-NEDE}). Crucially, when combining all data sets, the $Q_{\rm dmap}$ tension for $H_0$ and $S_8$, are $0.4\sigma$ and $2.2\sigma$ respectively, showing NEDE's ability to resolve both tensions. At the same time, the $S_8$ Gaussian tension remains at $1.9\sigma$ independently of whether the $H_0$ prior is added. This reflects one of the main points of the present article: as opposed to other EDE-like models, the presence of the NEDE fluid ($f_\mathrm{NEDE} > 0$)  does not exacerbate the $S_8$ tension for $\Omega_\phi > 0$. Moreover, looking at individual $\chi^2$ improvements in the analysis with all data sets considered, NEDE is compatible with all observations.

When comparing the goodness of fit via $\chi^2$ values, NEDE gives a better fit when compared to LCDM by a few $\chi^2$ units whenever there is no $H_0$ prior included in the analysis and by 20 or more units for simulations including the S$H_0$ES prior. The latter simulations are also better fits even when compared to the restricted NEDE model with $\Omega_\phi = 0$, showing that the trigger perturbations' evolution does not only alleviate the $S_8$ tension but also allows the overall model to give a better fit.

\begin{table}[hbtp]
	\centering
	\begin{tabular}{>{\arraybackslash}p{4cm}|>{\centering\arraybackslash}p{2.5cm}|>{\centering\arraybackslash}p{2.5cm}|>{\centering\arraybackslash}p{2.5cm}|>{\centering\arraybackslash}p{2.5cm}}
		\centering NEDE $(\Omega_\phi > 0)$           & Baseline                    & + LSS        & + $H_0$      & + $H_0$ + LSS \\ \hline\hline
		Pl.2018 low $\ell$ TT                            & -2.1                        & -1.7         & -1.5         & -1.0          \\ \hline
		Pl.2018 low $\ell$ EE                            & 0.4                         & -1.4         & -0.9         & 0.2           \\ \hline
		Pl.2018 high $\ell$ TTTEEE                       & -1.3                        & 2.0          & -5.2         & -1.6          \\ \hline
		Pl.2018 Lensing                                  & 0.8                         & 0.4          & 0.5          & 0.3           \\ \hline
		BAO SDSS DR7 mgs                                 & 0.8                         & -0.2         & -0.5         & -0.7          \\ \hline
		BAO 6dF 2011                                     & -0.1                        & 0.0          & 0.0          & 0.0           \\ \hline
		BAO SDSS DR12                                    & -1.7                        & 0.3          & 0.0          & 0.2           \\ \hline
		SN Pantheon                                      & -0.4                        & 0.0          & 0.1          & 0.1           \\ \hline
		S$H_0$ES 2021                                    & --                          & --           & -16.4        & -16.5         \\ \hline
		LSS                                              & --                          & -1.8         & --           & -1.2          \\ \hline\hline
		\centering $\Delta\, \chi^2$                     & -3.7                        & -2.4         & -23.9        & -20.1         \\ \hline
		\centering $H_0-Q_{\rm dmap}$ Tension            & --                          & --           & 1.41         & 0.38          \\ \hline
		\centering $S_8-Q_{\rm dmap}$ Tension            & --                          & 2.55$\,{}^\lozenge$         & --           & 2.15          \\ \hline
		\centering $H_0$ Gauss.Tension${}^\blacklozenge$ & (1.79$\sigma$) 3.05$\sigma$ & 3.11$\sigma$ & --           & --            \\ \hline
		\centering $S_8$ Gauss.Tension${}^\blacklozenge$ & (1.96$\sigma$) 1.88$\sigma$ & --           & 1.85$\sigma$ & --            \\ \hline
	\end{tabular}\\
	\begin{flushleft}
	{\footnotesize ${}^\lozenge$ This value corresponds to the best-fit of an MCMC chain performed while keeping $w\nede$ and $z_*$ fixed, it is the best result obtained via Cobaya's minimization procedure, however we expect the real best-fit might differ.\\
	${}^\blacklozenge$ In parenthesis, we report the Gaussian tension computed with the posteriors of the simulation with fixed $z_*$ and $w\nede$, to account for sampling volume effects (see Table~\ref{tab:paramsFixRun} and Table~\ref{tab:chiSqNEDEfxParams} in Appendix~\ref{app:tables})}.
	\end{flushleft}
	\vspace*{-.5cm}
	\caption{Table of $\Delta\chi^2= \chi^2\nede - \chi^2_{\Lambda{\rm CDM}}$ differences between the NEDE model and $\Lambda$CDM aggregated by datasets.}
	\label{tab:chiSqDiffNEDEvsLCDM}
\end{table}

\begin{figure}[t]
	\centering
	\includegraphics[clip,width=\textwidth, trim=0 0 0 0]{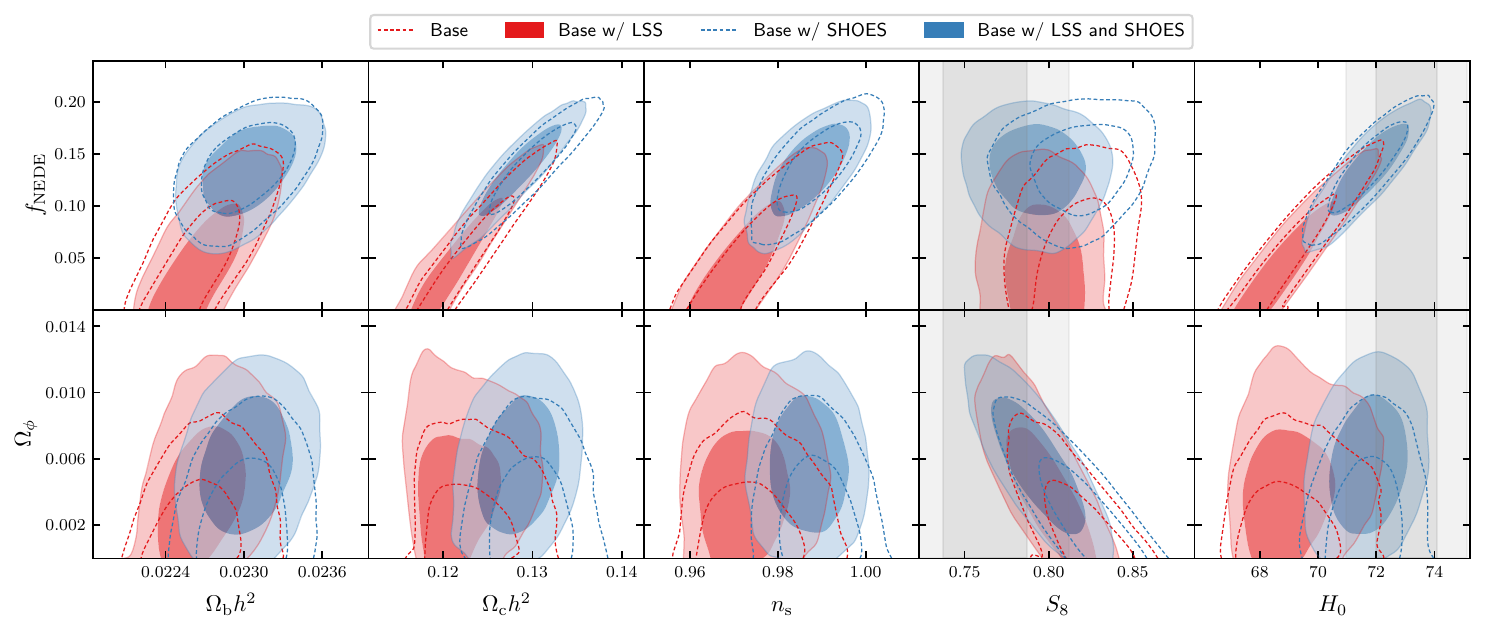}
	\caption{Contour comparison of the parameters $f\nede$ and $\Omega_{\phi}$ for the MCMC chains for the NEDE model with different data sets. The vertical bands correspond to the $1\sigma$ and $2\sigma$ standard deviations of the constraint on $H_0$ from the S$H_0$ES collaboration~\cite{Riess2022} and the one on $S_8$ from the combined analysis of KiDS+Viking+DES~\cite{Joudaki2020}.}
	\label{fig:f-OmegaXLCDM}
\end{figure}

The different $68 \%$ and $95 \%$ C.L. contours for the NEDE model, pertaining to the parameters $f\nede$ and $\Omega_\phi$ can be appreciated in Fig.~\ref{fig:f-OmegaXLCDM}. Here, it is worth noticing how $f\nede$ correlates positively with $H_0$ while $\Omega_\phi$ is anti-correlated with $S_8$. It is also interesting to witness how the inclusion of a prior for $H_0$ and $S_8$ leads to more than $5\sigma$ evidence for $f_\mathrm{NEDE}$. Additionally, we observe that while $f\nede$ correlates positively with all parameters shown in this figure, except for $S_8$ as is known, $\Omega_\phi$ has close to no effect on  the parameters that correlate with $f\nede$, anti-correlates with $S_8$ and has a mild positive correlation with $H_0$. In other words, the trigger acts mostly independent of the NEDE mechanism while slightly assisting it.

We now discuss the best-fit, posterior means, and errors for all MCMC samples, which are collected and presented in Table~\ref{tab:paramResults}. The first seven rows of values belong to the $\Lambda$CDM model. Of particular relevance for the present discussion are  $H_0$ and $S_8$, which, independent of the data set combination used, remain in tension. To be specific, we obtain a mean for $H_0$ that ranges between $67.7 - 68.6\kmsMpc$ with standard deviations below 0.4 $\kmsMpc$. This is compatible with the standard literature and corresponds to a Hubble tension of order $5 \sigma$ when compared with SH$_0$ES. A secondary but similar behavior is seen when looking at the $S_8$ parameter. The model predicts $S_8 = 0.824 \pm 0.01$ for the baseline analysis, which amounts to a $2.3 \sigma$ $S_8$ tension when compared with our LSS prior.

\begin{table}[!htbp]
	\centering
	\renewcommand{\arraystretch}{1.2}
	\setlength\tabcolsep{1pt}
	\fontsize{8}{9.5}\selectfont
	\begin{tabular}{>{\centering\arraybackslash}p{2.15cm}||>{\centering\arraybackslash}p{3.5cm}|>{\centering\arraybackslash}p{3.5cm}|>{\centering\arraybackslash}p{3.5cm}|>{\centering\arraybackslash}p{3.5cm}}
		                                           & Baseline                                  & + LSS                                     & + S$H_0$ES                                & + S$H_0$ES + LSS                          \\ \hline\hline
		$\Lambda$CDM                               & (Bestfit) Mean$_{\rm Lower}^{\rm Upper}$ & (Bestfit) Mean$^{\rm Upper}_{\rm Lower}$ & (Bestfit) Mean$^{\rm Upper}_{\rm Lower}$ & (Bestfit) Mean$_{\rm Lower}^{\rm Upper}$ \\ \hline\hline
		$\Omega_\mathrm{b} h^2$                    & $(0.022)\; 0.02242\pm 0.00013$            & $(0.022)\; 0.02260\pm 0.00013$            & $(0.023)\; 0.02257\pm 0.00013$            & $(0.023)\; 0.02260\pm 0.00013$            \\ \hline
		$\Omega_\mathrm{c} h^2$                    & $(0.120)\; 0.11921\pm 0.00091$            & $(0.118)\; 0.11730\pm 0.00083$            & $(0.117)\; 0.11773\pm 0.00087$            & $(0.117)\; 0.11730\pm 0.00082$            \\ \hline
		$H_0\,${\footnotesize [km/s/Mpc]}          & $(67.446)\; 67.72\pm 0.41$                & $(68.181)\; 68.64\pm 0.38$                & $(68.636)\; 68.45\pm 0.39$                & $(68.647)\; 68.64\pm 0.37$                \\ \hline
		$\log(10^{10} A_\mathrm{s})$               & $(3.046)\; 3.048\pm 0.014$                & $(3.059)\; 3.050^{+0.014}_{-0.016}$       & $(3.057)\; 3.054^{+0.014}_{-0.016}$       & $(3.060)\; 3.051\pm 0.015$                \\ \hline
		$n_\mathrm{s}$                             & $(0.965)\; 0.9667\pm 0.0036$              & $(0.967)\; 0.9712\pm 0.0036$              & $(0.971)\; 0.9704\pm 0.0037$              & $(0.970)\; 0.9712\pm 0.0036$              \\ \hline
		$\tau_\mathrm{reio}$                       & $(0.053)\; 0.0566\pm 0.0072$              & $(0.060)\; 0.0597^{+0.0069}_{-0.0080}$    & $(0.060)\; 0.0611^{+0.0071}_{-0.0081}$    & $(0.055)\; 0.0600^{+0.0068}_{-0.0078}$    \\ \hline
		$S_8$                                      & $(0.830)\; 0.824\pm 0.010$                & $(0.816)\; 0.8030\pm 0.0092$              & $(0.806)\; 0.809\pm 0.010$                & $(0.806)\; 0.8032\pm 0.0092$              \\ \hline\hline
		NEDE $(\Omega_\phi = 0)$                   &                                           &                                           &                                           &                                           \\ \hline\hline
		$\Omega_\mathrm{b} h^2$                    & $(0.0230)\; 0.0226^{+0.0002}_{-0.0003}$   & $(0.0220)\; 0.0226^{+0.0002}_{-0.0002}$   & $(0.0230)\; 0.0230^{+0.0002}_{-0.0002}$   & $(0.0230)\; 0.0229^{+0.0002}_{-0.0002}$   \\ \hline
		$\Omega_\mathrm{c} h^2$                    & $(0.1240)\; 0.1228^{+0.0017}_{-0.0040}$   & $(0.1190)\; 0.1207^{+0.0011}_{-0.0027}$   & $(0.1320)\; 0.1298^{+0.0033}_{-0.0033}$   & $(0.1300)\; 0.1267^{+0.0030}_{-0.0030}$   \\ \hline
		$H_0\,${\footnotesize [km/s/Mpc]}          & $(69.53)\; 68.82^{+0.65}_{-1.30}$         & $(68.09)\; 68.68^{+0.50}_{-0.96}$         & $(71.76)\; 71.51^{+0.89}_{-0.89}$         & $(72.02)\; 71.25^{+0.88}_{-0.88}$         \\ \hline
		$\log(10^{10} A_\mathrm{s})$               & $(3.065)\; 3.054^{+0.015}_{-0.015}$       & $(3.047)\; 3.046^{+0.015}_{-0.015}$       & $(3.068)\; 3.069^{+0.015}_{-0.015}$       & $(3.067)\; 3.060^{+0.015}_{-0.015}$       \\ \hline
		$n_\mathrm{s}$                             & $(0.9770)\; 0.9733^{+0.0055}_{-0.0086}$   & $(0.9670)\; 0.9717^{+0.0046}_{-0.0068}$   & $(0.9910)\; 0.9897^{+0.0064}_{-0.0064}$   & $(0.9930)\; 0.9870^{+0.0064}_{-0.0064}$   \\ \hline
		$\tau_\mathrm{reio}$                       & $(0.0560)\; 0.0564^{+0.0073}_{-0.0073}$   & $(0.0540)\; 0.0550^{+0.0072}_{-0.0072}$   & $(0.0560)\; 0.0579^{+0.0069}_{-0.0078}$   & $(0.0580)\; 0.0566^{+0.0074}_{-0.0074}$   \\ \hline
		\rowcolor{gray!15} $f_\mathrm{NEDE}$       & $(0.0660)\; < 0.114 $                     & $(0.0050)\; < 0.0818$                     & $(0.1440)\; 0.1280^{+0.0290}_{-0.0290}$   & $(0.1330)\; 0.1050^{+0.0320}_{-0.0270}$   \\ \hline
		\rowcolor{gray!15} $3\omega_\mathrm{NEDE}$ & $(2.03)\; > 1.40$                         & $(1.48)\;$ Unconstrained                  & $(2.08)\; 2.12^{+0.12}_{-0.20}$           & $(2.07)\; 2.14^{+0.15}_{-0.26}$           \\ \hline
		\rowcolor{gray!15} $z_\mathrm{decay}$      & $(5016)\; 4534^{+2000}_{-2000}$           & $(6854)\; 4257^{+2000}_{-4000}$           & $(4879)\; 4615^{+800}_{-800}$             & $(4561)\; 4467^{+900}_{-900}$             \\ \hline
		$m \; [{\rm Mpc}^{-1}]^\blacklozenge$               & $(366)\; < 914$                           & $(621)\; < 956$                           & $(368)\; 336^{+80}_{-100}$                & $(323)\; 315^{+80}_{-100}$                \\ \hline
		$S_8$                                       & $(0.8350)\; 0.8310^{+0.0120}_{-0.0120}$   & $(0.8170)\; 0.8180^{+0.0100}_{-0.0100}$   & $(0.8460)\; 0.8400^{+0.0130}_{-0.0130}$   & $(0.8320)\; 0.8240^{+0.0110}_{-0.0110}$   \\ \hline\hline
		NEDE       $(\Omega_\phi > 0)$          &                                           &                                           &                                           &                                           \\ \hline\hline
		$\Omega_\mathrm{b} h^2$                    & $(0.0230)\; 0.0227^{+0.0002}_{-0.0003}$   & $(0.0230)\; 0.0227^{+0.0002}_{-0.0003}$   & $(0.0230)\; 0.0230^{+0.0002}_{-0.0002}$   & $(0.0230)\; 0.0230^{+0.0002}_{-0.0002}$   \\ \hline
		$\Omega_\mathrm{c} h^2$                    & $(0.1290)\; 0.1232^{+0.0023}_{-0.0044}$   & $(0.1250)\; 0.1219^{+0.0021}_{-0.0040}$   & $(0.1320)\; 0.1299^{+0.0032}_{-0.0032}$   & $(0.1300)\; 0.1284^{+0.0030}_{-0.0030}$   \\ \hline
		$H_0\,${\footnotesize [km/s/Mpc]}          & $(70.84)\; 69.06^{+0.78}_{-1.40}$         & $(70.26)\; 69.10^{+0.72}_{-1.40}$         & $(71.76)\; 71.69^{+0.89}_{-0.89}$         & $(71.74)\; 71.71^{+0.88}_{-0.88}$         \\ \hline
		$\log(10^{10} A_\mathrm{s})$               & $(3.0670)\; 3.0570^{+0.0160}_{-0.0160}$   & $(3.0630)\; 3.0540^{+0.0160}_{-0.0160}$   & $(3.0680)\; 3.0720^{+0.0150}_{-0.0150}$   & $(3.0690)\; 3.0700^{+0.0160}_{-0.0160}$   \\ \hline
		$n_\mathrm{s}$                             & $(0.9840)\; 0.9739^{+0.0060}_{-0.0090}$   & $(0.9810)\; 0.9729^{+0.0057}_{-0.0082}$   & $(0.9910)\; 0.9889^{+0.0062}_{-0.0062}$   & $(0.9880)\; 0.9868^{+0.0058}_{-0.0058}$   \\ \hline
		$\tau_\mathrm{reio}$                       & $(0.0550)\; 0.0572^{+0.0074}_{-0.0074}$   & $(0.0550)\; 0.0569^{+0.0074}_{-0.0074}$   & $(0.0560)\; 0.0591^{+0.0070}_{-0.0080}$   & $(0.0560)\; 0.0594^{+0.0070}_{-0.0080}$   \\ \hline
		\rowcolor{gray!15} $f_\mathrm{NEDE}$       & $(0.1140)\; < 0.130$                      & $(0.0900)\; < 0.125$                      & $(0.1440)\; 0.1370^{+0.0300}_{-0.0260}$   & $(0.1460)\; 0.1340^{+0.0320}_{-0.0250}$   \\ \hline
		\rowcolor{gray!15} $z_\mathrm{decay}$      & $(4441)\; 4911^{+1000}_{-2000}$           & $(4499)\; 5084^{+1000}_{-3000}$           & $(4879)\; 4535^{+600}_{-700}$             & $(4626)\; 4414^{+500}_{-800}$             \\ \hline
		\rowcolor{gray!15} $3\omega_\mathrm{NEDE}$ & $(2.09)\; > 1.41$                         & $(2.06)\; 2.08^{+0.36}_{-0.53}$           & $(2.08)\; 2.09^{+0.11}_{-0.19}$           & $(2.03)\; 2.05^{+0.12}_{-0.18}$           \\ \hline
		\rowcolor{gray!15} $\Omega_{0,\phi}$       & $(0.0010)\; < 0.0068$                     & $(0.0040)\; < 0.00988$                    & $(0.0000)\; < 0.00776$                    & $(0.0060)\; 0.0057^{+0.0025}_{-0.0029}$   \\ \hline
		$m \; [{\rm Mpc}^{-1}]^\blacklozenge$               & $(304.88)\; < 1010$                       & $(306.88)\; 418.00^{+60.00}_{-300.00}$    & $(368.36)\; 329.00^{+70.00}_{-100.00}$    & $(336.11)\; 314.00^{+50.00}_{-100.00}$    \\ \hline
		$\phi_{\rm ini}\; [{M}_{\rm Pl}]$          & $(0.0530)\; 0.0670^{+0.0280}_{-0.0350}$   & $(0.0920)\; 0.0920^{+0.0350}_{-0.0280}$   & $(0.0000)\; 0.0800^{+0.0340}_{-0.0340}$   & $(0.1170)\; 0.1130^{+0.0330}_{-0.0230}$   \\ \hline
		$S_8$                                      & $(0.8330)\; 0.8160^{+0.0180}_{-0.0150}$   & $(0.8040)\; 0.7970^{+0.0180}_{-0.0150}$   & $(0.8460)\; 0.8180^{+0.0230}_{-0.0170}$   & $(0.7990)\; 0.7930^{+0.0180}_{-0.0180}$   \\ \hline\hline
	\end{tabular}\\
	\vspace*{-.2cm}
		\begin{flushleft}
		{\footnotesize ${}^\blacklozenge$ For reference $313\, {\rm Mpc}^{-1} \approx 2.0 \times 10^{-27}\, {\rm eV}$ .}
		\end{flushleft}
	\vspace*{-.8cm}

	\caption{\small All cosmological parameters for different models and combinations of datasets. The NEDE parameters are shaded and additional derived parameters are appended below them. The $1\sigma$ standard deviation is shown except for one-sided constraints, which correspond to the $2\sigma$ range. }
	\label{tab:paramResults}
\end{table}

The best-fit, posterior means, and errors for the restricted NEDE model with $(\Omega_\phi = 0 )$ are presented immediately below those for $\Lambda$CDM in Table~\ref{tab:paramResults}. Highlighted in gray are the rows belonging to the unique model parameters, below which we present the derived parameters $m$ and $S_8$. This is an update on values already reported in previous studies~\cite{Niedermann:2020dwg, Cruz:2022oqk, Cruz:2023cxy}. All simulations sampled the three parameters $(f\nede, w\nede,\log_{10}(z_*))$. The equation of state for the NEDE fluid is seen to be relatively unaffected by the data set combination, giving a $1 \sigma$ range of $3w\nede = 1.9 - 2.3$, provided the simulations included the $H_0$ prior. The standard deviation is significantly larger for the baseline and baseline + LSS analysis, which appear less constraining due to the volume effects discussed before. Very similarly, the decay redshift, $z_*$, is consistently around 4500 with a large standard deviation in the first two columns, which again is affected by the sampling volume enlargement, and with halved errors when the $H_0$ prior is included. Regarding the fraction of NEDE, we observe results consistent with previous analyses, where we obtain a strict one-sided bound when including LSS or a relaxed bound for the baseline case, provided neither $z_*$ nor $w\nede$ was fixed (see Table~\ref{tab:paramsFixRun} and Fig.~\ref{fig:nedeTriangleFixed} for an analysis that avoids the sampling volume issues). The simulation, including all data sets (last column), had not been considered previously and shows a fraction of NEDE at decay $z_*$ of $f\nede = 0.105^{+0.032}_{-0.027}$. We attribute the decrease in average fraction, when compared to the run without LSS, to the pull exerted by the LSS likelihood. This last point is connected to the $S_8$ row, which shows an increase across the table when compared to the $\Lambda$CDM counterparts, making the $S_8$ tension slightly worse. The $S_8$ parameter, for this model, gives a posterior of $S_8 = 0.831\pm 0.012$, for the run that uses only the baseline and is positively correlated with the amount of NEDE in the given run. In particular, the $S_8$ tension is still significant with more than $2.4 \sigma$.

The results for the NEDE $(\Omega_{\phi} > 0)$ model are tabulated in the bottom rows of Table~\ref{tab:paramResults}. As before, rows highlighted in gray belong specifically to the unique NEDE parameters. These rows illustrate the main message of this paper, which, put shortly, is that the average value for the $S_8$ parameter is generally reduced when compared to both $\Lambda$CDM or the restricted NEDE. Using all compatible data sets, i.e.~baseline+$H_0$+LSS, the posterior obtained is $S_8 = 0.793\pm 0.018$, which reduces the tension with the weak lensing data prior to $2.15\sigma$, in the $Q_{\rm dmap}$ sense. It is important to notice, this is done without affecting NEDE's ability to resolve the Hubble tension, giving $H_0=71.71\pm 0.88\kmsMpc$. The baseline+$H_0$, as well as the baseline + $H_0$ + LSS results, show a preference for $f\nede>0$ of more than $5\sigma$. Moreover, sampling over $\Omega_{\phi}$ has not enlarged their standard deviations, to the contrary, the lower tail is made narrower increasing the evidence for NEDE. As a final remark, we note that the conclusion that NEDE leads to a bluer primordial scalar spectrum, larger $n_s$, remains true, and we confirm that the simplest curvaton model~\cite{Enqvist:2001zp,Lyth:2001nq,Moroi:2001ct} is preferred over Starobinsky inflation~\cite{Starobinsky:1980te} in the NEDE framework \cite{Cruz:2022oqk}.

\section{Conclusions}
\label{sec:conclusion}

In this work, we have demonstrated that a triggered dark energy phase transition, as proposed by the Cold NEDE model, has the potential to simultaneously address two major tensions in cosmology. The $H_0$ tension is resolved through an energy injection around matter-radiation equality, which is characteristic of EDE-type models. The $S_8$ tension, on the other hand, is relieved due to the presence of coherent oscillations of the ultralight trigger field already present in Cold NEDE. To be specific, after the phase transition, the trigger field acts as a small non-thermal contribution to dark matter, which dampens the power spectrum on small scales. It is a vital ingredient in the Cold NEDE model because it synchronizes the phase transition, sources dark sector fluid perturbations, and prevents the occurrence of early vacuum bubbles that could introduce additional, unobserved anisotropies in the CMB. The only difference with our previous work in \cite{Niedermann:2020dwg} is that we drop the assumption that the trigger field makes a negligible contribution to the energy budget.  In other words, no new features are introduced at the fundamental level. Instead, we scan the trigger dark matter abundance $\Omega_\phi$ as an additional parameter in our phenomenological description (rather than assuming $\Omega_\phi \ll1$). The phenomenological potential of an ultralight field in combination with NEDE has been pointed out before by Allali, Hertzberg, and Rompineve~\cite{Allali:2021azp}. There, however, the ultralight scalar was an added component to the NEDE fields , and its mass could be chosen freely. Interestingly, despite being more constrained, our minimal scenario achieves a similar level of improvement with regard to both tensions.

As our main result, we find that both DMAP tensions are reduced to almost below $2 \sigma$ when allowing for a sizable amount of $\Omega_\phi$. At the same time, we maintain an excellent fit to CMB, supernovae, and BAO data. A second crucial observation is that allowing $\Omega_\phi$ to vary reduces the error bars on $f_\mathrm{NEDE}$, explicitly $f_\mathrm{NEDE} = 13.4^{+3.2}_{-2.5} \, \% $, and thus increases the Gaussian evidence above $5 \sigma$ when including a SH$_0$ES prior on $H_0$. This has to be contrasted with $\Omega_\phi = 0$, where we find a weaker $3.9 \sigma$ and $4.4 \sigma$ evidence with and without additional LSS data, respectively. It is also remarkable because normally, error bars increase as the number of free parameters is increased. Finally, additional LSS data assists the mechanism that solves the $H_0$ tension. This means that one tension is not solved at the price of making the other tension worse; instead, both mechanisms ``pull'' in the same direction.  For example, when including additional LSS data, the $H_0$ DMAP tension is reduced from $1.3 \sigma$ to $0.4 \sigma$, completely eradicating the $H_0$ tension. Again, this should be contrasted with previous studies of NEDE with  $\Omega_\phi=0$ (and other EDE type models), where the converse was true, and LSS data hampered the model's potential to resolve the $H_0$ tension (albeit weakly~\cite{Niedermann:2020qbw}).

While this work had its focus on investigating the phenomenological potential of Cold NEDE as a simultaneous solution to both the $H_0$ and $S_8$ tension, it also exemplifies the importance of having a microscopic theory that guides any effort to modify the $\Lambda$CDM model in a physically motivated way. After all, the trigger field was not introduced to address the $S_8$ tension but to make a first-order phase transition a viable scenario for a new phase of dark energy. We stress, however, that different model-building aspects require further investigation. For example, the fluid is currently treated phenomenologically, despite the fact that, ultimately, it will be constrained by the microscopic parameters characterizing the tunneling potential. In addition, our findings also reinforce the idea that, within Cold NEDE, ultralight physics needs to be invoked to trigger the phase transition and restore cosmological concordance. This, in turn, raises the question of how to explain the origin of these tiny mass scales.  As a preliminary step, we have outlined a possible embedding of Cold NEDE in a two-axion framework. Here the small masses of both, the trigger and tunneling field, are protected by approximate shift symmetries, and the small coupling between both fields can be understood as a consequence of the hierarchy between the axion decay constant and the breaking scale of a global shift symmetry. In the next step, it will therefore be important to embed Cold NEDE in a complete axion framework. In such a framework, it will be interesting to revisit phenomenological constraints on axion physics, such as cosmic birefringence \cite{Nakatsuka:2022epj,Eskilt:2023nxm}.

To summarize, our findings establish the NEDE framework as a theoretically well-motivated, minimal, and phenomenologically promising framework for solving the present tensions in cosmology.

\section*{Acknowledgments}
The authors thank Vivian Poulin for comments on the draft. J.S.C. and M.S.S. are supported by Independent Research Fund Denmark grant 0135-00378B. The work of F.N. was supported by VR Starting Grant 2022-03160 of the Swedish Research Council.

\appendix
\section{Fluid matching} \label{app:matching}

Here, we describe how we match the trigger field $\phi(x)$ in the Cold NEDE model to an effective fluid description at time $t_m > t_*$. This is necessary if we want to keep track of the trigger component after the phase transition and infer the amount of dark matter today. To that end, we use exactly the scheme proposed in \cite{Passaglia:2022bcr}, which has also been implemented in the newest release, \textsc{TriggerCLASSv6.1}.

\begin{figure}[!htbp]
	\centering
	\includegraphics[clip, width=\textwidth]{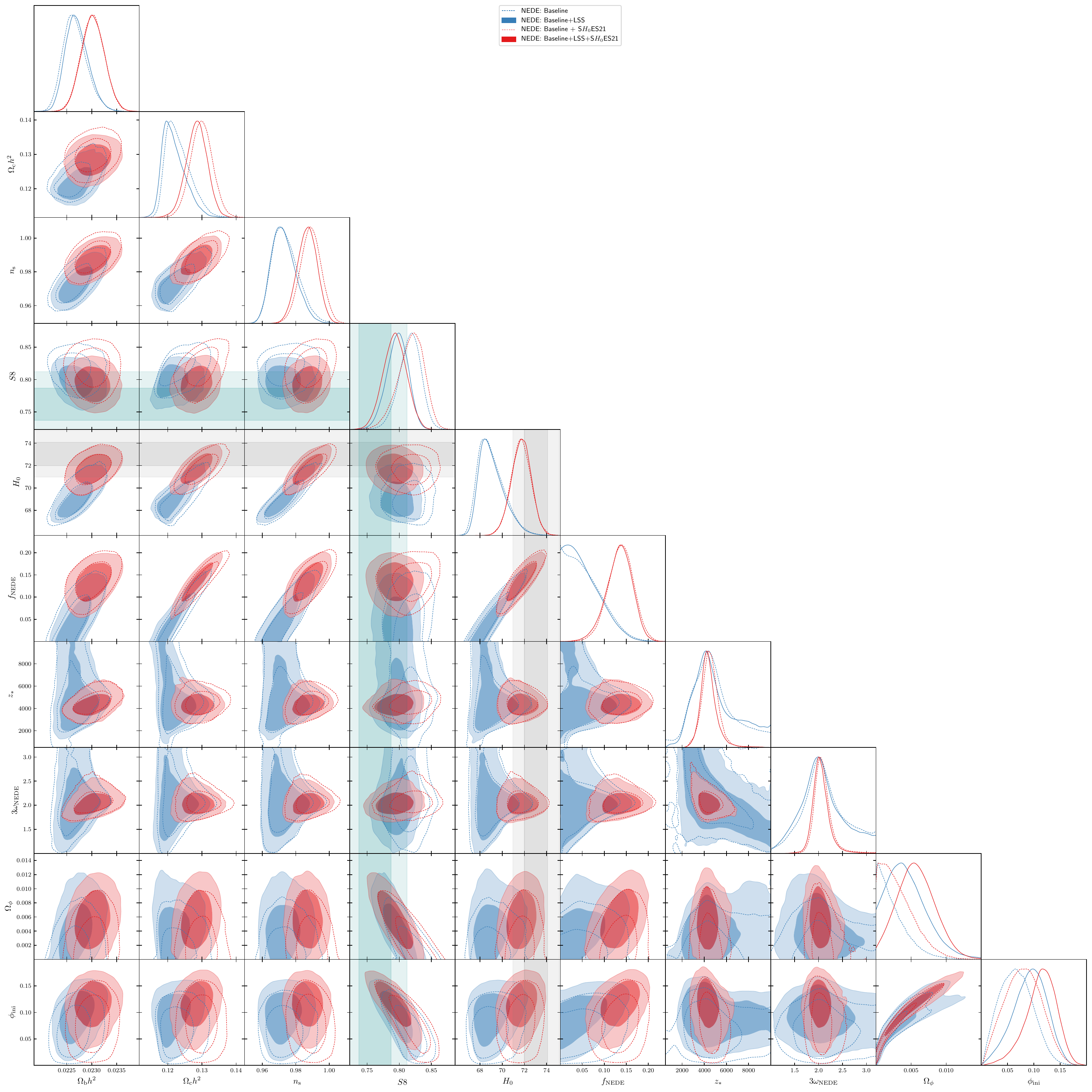}
	\caption{Triangle plot showing the $68 \%$ and $95 \%$ C.L. contours  for some of the relevant parameters of the model with $\Omega_\phi >0$.}
	\label{fig:nedeTriangle}
\end{figure}

The strategy consists of parameterize the field and its perturbations in such a way that oscillations are factored out. From Eq.~\eqref{eq:bkgTriggerEOM} it is clear that the asymptotic period of the background trigger oscillations is of the order of $1/m$. This makes it natural to introduce $\bar{t} \equiv m t$ with $t$ the cosmological time. We denote derivatives with respect to $\bar{t}$ as a dot. Introducing the fields $\phi_s$ and $\phi_c$, as well as $\delta \phi_s$ and $\delta \phi_c$, such that
\begin{subequations}
	\begin{align}
 \phi(\bar{t}) &= \phi_c (\bar{t})\cos[\bar{t} - \bar{t}_m] + \phi_s(\bar{t})\sin[\bar{t}-\bar{t}_m],\\
 \delta\phi(\bar{t}) &= \delta\phi_c (\bar{t})\cos[\bar{t} - \bar{t}_m] + \delta\phi_s(\bar{t})\sin[\bar{t}-\bar{t}_m],
\end{align}
\label{eq:triggerReparam}
\end{subequations}
it is possible to construct an effective fluid description that can be evolved {efficiently with controlled numerical error. The fluid quantities are then obtained after inserting Eqs.~\eqref{eq:triggerReparam} into the energy-momentum tensor for a scalar field and integrating over a period, amounting to the replacement $\sin^2(\cdot) \to 1/2$, $ \cos^2(\cdot) \to 1/2 $ and $\cos(\cdot) \sin(\cdot) \to 0$,
\begin{subequations}
	\begin{align}
		\rho^{\rm efa}_\phi\big|_m &= \frac{1}{2}m^2 \left(\phi_c^2 + \phi_s^2 + \frac{\dot \phi^2_c}{2} + \frac{\dot \phi^2_s}{2} - \phi_c \dot \phi_s + \phi_s \dot \phi_c \right)_m,\\[4pt]
		p^{\rm efa}_\phi\big|_m &= \frac{1}{2}m^2 \left(\frac{\dot \phi^2_c}{2} + \frac{\dot \phi^2_s}{2} - \phi_c \dot \phi_s + \phi_s \dot \phi_c \right)_m.
	\end{align}
\label{eq:scalarToEF}
\end{subequations}
Similarly, for the energy, pressure and velocity perturbations,
\begin{subequations}
	\begin{align}
		\delta\rho^{\rm efa}_\phi\big|_m &= \frac{1}{2}m^2 \left[\phi_s\delta \dot \phi_c -\phi_c\delta \dot \phi_s + \delta \dot \phi_c \dot \phi_c + \delta \dot \phi_s \dot \phi_s + \delta \phi_s(2\phi_s + \dot \phi_c) + \delta\phi_c(2\phi_c - \dot \phi_s) \right]_m,\\[4pt]
		\delta p^{\rm efa}_\phi\big|_m &= \delta\rho^{\rm ef}_\phi -m^2(\delta\phi_s \phi_s + \delta\phi_c \phi_c),\\[4pt]
		(\rho^{\rm efa}_\phi + p^{\rm ef}_\phi)\theta^{\rm ef}_\phi\big|_m &= \frac{k^2 m}{2 a}\left( \delta\phi_c(\phi_s + \dot \phi_c) + \delta\phi_s(-\phi_c + \dot \phi_s)\right).
	\end{align}
\label{eq:efPertQuantities}
\end{subequations}
The new variables on the right hand side of \eqref{eq:scalarToEF} and \eqref{eq:efPertQuantities} are determined by using \eqref{eq:triggerReparam} to match the field value and its time derivative at time $\bar{t}_m$. As this description doubles the number of fields, two constraints can be imposed on $\dot \phi_s$ and $\dot \phi_c$, and two more on their perturbations $\delta\dot \phi_s$ and $\delta \dot \phi_c$. A suitable choice that suppresses fast oscillatory modes, and hence reduces the matching error, has been derived in~\cite{Passaglia:2022bcr}; explicitly,
\begin{subequations}
	\begin{align}
		\frac{\ddot \phi_{c,s}}{\dot \phi_{c,s}}\bigg|_m &= - \frac{1}{2}\frac{H}{m}\left(3 -2\frac{m}{H^2}\frac{{\rm d}H}{{\rm d}\bar{t}}\right)\bigg|_m,\\[4pt]
		\frac{ \delta \ddot\phi_{c,s}}{\delta\dot \phi_{c,s}}\bigg|_m &= - \frac{1}{2}\frac{H}{m}\left(3 -2\frac{m}{H^2}\frac{{\rm d}H}{{\rm d}\bar{t}}\right)\bigg|_m,
	\end{align}
\label{eq:matchTrigger}
\end{subequations}
where $\ddot \phi_{c,s}$ are given by
\begin{subequations}
\begin{align}
\ddot \phi_c &= -2 \dot \phi_s -3 \frac{H}{m} \left( \phi_s + \dot \phi_c \right)\,,\\
\ddot \phi_s &= 2 \dot \phi_c -3 \frac{H}{m} \left( -\phi_c + \dot \phi_s \right)\,,
\end{align}
\end{subequations}
and $\delta \ddot\phi_{c,s}$ by
\begin{subequations}
\begin{align}
\delta \ddot  \phi_c &= -2 \delta \dot \phi_s -3 \frac{H}{m} \left( \delta \phi_s +\delta \dot \phi_c \right) - \frac{k^2}{a^2 m^2} \delta \phi_c - \frac{\dot h}{2} \left(\delta \phi_s + \delta \dot \phi_c \right)\,,\\
\delta \ddot  \phi_s &= 2 \delta \dot \phi_c -3 \frac{H}{m} \left( -\delta \phi_c +\delta \dot \phi_s \right) - \frac{k^2}{a^2 m^2} \delta \phi_s - \frac{\dot h}{2} \left(-\delta \phi_c + \delta \dot \phi_s \right)\,.
\end{align}
\end{subequations}

The above constraints achieve an error suppression at background level of $\mathcal{O}(H_*/m)^3$ (and slightly milder suppression on perturbation level), which is sufficient for our current application.

\section{Triangle plots for NEDE}
\label{app:triangles}

In this section we include additional triangle plots, for all the MCMC chains computed for the different NEDE runs and data sets mentioned in the main text. We include the cosmological parameters that are relevant for the present discussion including the NEDE specific parameters.
\begin{figure}[tbp]
	\centering
	\includegraphics[clip, width=\textwidth]{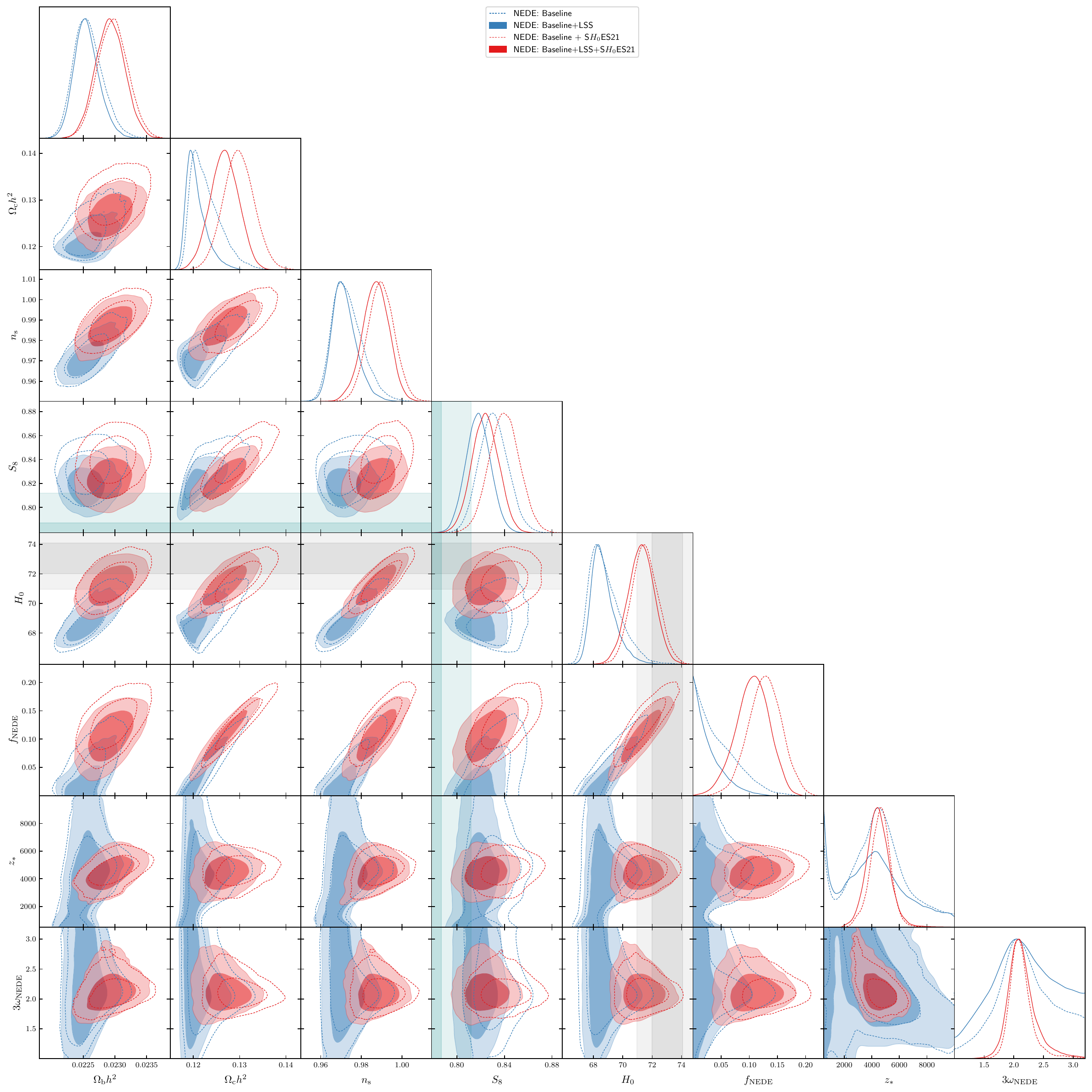}
	\caption{Triangle plot showing the $68 \%$ and $95 \%$ C.L. contours for some of the relevant parameters of the NEDE model with $\Omega_\phi = 0$.}
	\label{fig:nedeOmegaPhi0Triangle}
\end{figure}

Figure~\ref{fig:nedeOmegaPhi0Triangle} shows an overlay of the parameter contours for the restricted NEDE model with $\Omega_\phi= 0$, along with the $1 \sigma$ and $2\sigma$ bands of the likelihoods in tension, i.e.~$S_8$ and $H_0$. Besides the comments already given in the main body of the text, we can observe the effects on $n_s$ and the regime for $\phi_{\rm ini}$, where $n_s$ reaches values as high as 0.99 (at $68\%$ C.L.), favoring a near scale-invariant spectrum of perturbations, as most EDE-like models. This shows how adding the LSS data, does not have an impact on this parameter. In this figure, it is also possible to see how this model is unable to reach lower $S_8$ values than 0.81 at $68\%$ C.L., no matter what combination of data sets is used.

The corresponding triangle plot for the NEDE MCMC chains is shown in Fig.~\ref{fig:nedeTriangle}. The $S_8$ plots differ from the ones in Fig.~\ref{fig:nedeOmegaPhi0Triangle}, because here contours at $68\%$ C.L. overlap the $1\sigma$ band from the LSS measurements. Posteriors in this plot are Gaussian in a good approximation, and deviate only whenever the parameter space volume is enlarged (which  happens in the limit where $f_\mathrm{NEDE} \to 0$)}. Another, secondary, observation, is that considering the trigger’s full evolution seems to also lift some of the degeneracy in the $3w\nede - z_*$ plane, in such a way that the $z_*$ posterior for runs without an $H_0$ prior, although not fully Gaussian,  suffer from the volume effects of Fig.~\ref{fig:nedeOmegaPhi0Triangle} to a lesser extent. The values for $\phi_{\rm ini}$ are also presented in this figure, where it is possible to appreciate that no trans-Planckian field values are required.

\newpage

\begin{figure}[!htbp]
	\centering
	\includegraphics[clip, width=\textwidth]{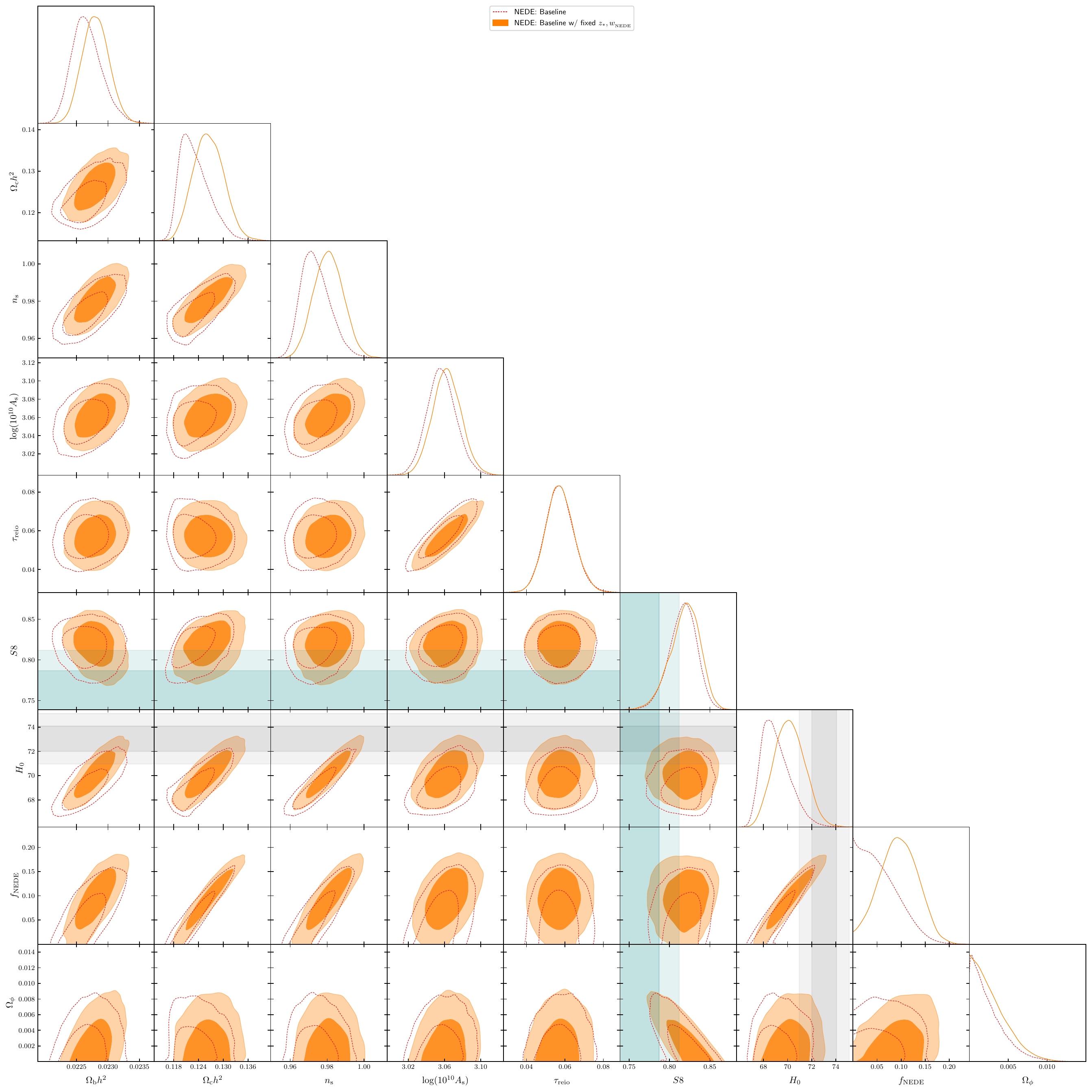}
	\caption{Triangle plot showing the $68 \%$ and $95 \%$ C.L. contours  for the NEDE model with $\Omega_\phi>0$ where we also fix $z_*= 4436$ and $3 w\nede= 2.088$ to account for sampling volume artifacts. }
	\label{fig:nedeTriangleFixed}
\end{figure}

The triangle plot in Fig.~\ref{fig:nedeTriangleFixed} depicts the contours produced by the NEDE model with $\Omega_phi>0$ when the parameters $z_*$ and $w\nede$ are fixed to their best-fit values obtained from a baseline run. As shown in \cite{Niedermann:2020dwg} and studied further in \cite{Cruz:2023cxy}, this procedure avoids the volume effects that otherwise occur in the limit $f_\mathrm{NEDE} \to 0$, allowing for an improved reading of the standard deviations. For example, this can  be seen by the change in the $f\nede$ posterior towards its left edge. We use these runs mainly to infer the Gaussian tension measure reported in Table~\ref{tab:chiSqDiffNEDEvsLCDM}.
\newpage

\section{Additional tables}
\label{app:tables}

\begin{table}
	\centering
	\renewcommand{\arraystretch}{1.1}
	\setlength\tabcolsep{2pt}
	\fontsize{8}{11}\selectfont
	\begin{tabular}{>{\arraybackslash\centering}p{3.5cm}|>{\centering\arraybackslash} p{4.5cm}}
		Parameter                       & (Bestfit)$\;$ Mean$^{\rm Upper}_{\rm Lower}$ \\ \hline\hline
		$\Omega_\mathrm{b} h^2$         & $(0.023)\; 0.02280\pm 0.00021$                \\ \hline
		$\Omega_\mathrm{c} h^2$         & $(0.128)\; 0.1265\pm 0.0036$                  \\ \hline
		$H_0$ [km/s/Mpc]                 & $(70.784)\; 70.2\pm 1.2$                      \\ \hline
		$\log(10^{10} A_\mathrm{s})$    & $(3.073)\; 3.063\pm 0.016$                    \\ \hline
		$n_\mathrm{s}$                  & $(0.986)\; 0.9808\pm 0.0078$                  \\ \hline
		$\tau_\mathrm{reio}$            & $(0.056)\; 0.0574\pm 0.0073$                  \\ \hline
		$f_\mathrm{NEDE}$               & $(0.109)\; 0.094\pm 0.039$                    \\ \hline
		$\log_{10}(m)$             & $(2.482)\; 2.479\pm 0.012$                    \\ \hline
		$m\, [{\rm Mpc}^{-1}] $    & $(303.625)\; 301.3\pm 8.3$                    \\ \hline
		$\phi_{\rm ini}\, [M_{\rm Pl}]$ & $(0.023)\; 0.073\pm 0.032$                    \\ \hline
		$\Omega_{\phi}$                 & $(0.000)\; < 0.00351$                         \\ \hline
		$S_8$                           & $(0.841)\; 0.819^{+0.021}_{-0.015}$           \\ \hline
	\end{tabular}
\caption{Bestfit, posterior means and error for the MCMC simulation of NEDE with $\Omega_\phi>0$ and fixed $z_* = 4436$ and $3w\nede = 2.088$.}
\label{tab:paramsFixRun}
\end{table}

\begin{table}
	\centering
	\renewcommand{\arraystretch}{1.1}
	\setlength\tabcolsep{2pt}
	\fontsize{8}{11}\selectfont
	\begin{tabular}{>{\arraybackslash} p{3.5cm}|>{\centering\arraybackslash}p{2.5cm}}
	NEDE fixed $z_*,w\nede$ & $\chi^2$\\ \hline \hline
	 Pl.2018 low $\ell$ TT        &	21.09   \\ \hline
	 Pl.2018 low $\ell$ EE        &	396.32  \\ \hline
	 Pl.2018 high $\ell$ TTTEEE   &	2336.86 \\ \hline
	 Pl.2018 Lensing              &	9.60    \\ \hline
	 BAO SDSS DR7 mgs             &	1.78    \\ \hline
	 BAO 6dF 2011                 &	0.00    \\ \hline
	 BAO SDSS DR12                &	3.52    \\ \hline
	 SN Pantheon                  &	1034.77 \\ \hline\hline
	 \centering $\Delta\, \chi^2$ &	3803.94 \\ \hline
	\end{tabular}
	\caption{Individual $\chi^2$ per likelihood for the MCMC chains of the NEDE model with $\Omega_\phi>0$ and fixed parameters $z_*=4436$ and $3w\nede = 2.088$.}
	\label{tab:chiSqNEDEfxParams}
\end{table}

\begin{table}[!htbp]
	\centering
	\renewcommand{\arraystretch}{1.1}
	\setlength\tabcolsep{2pt}
	\fontsize{8}{11}\selectfont
	\begin{tabular}{>{\arraybackslash}p{3.5cm}|>{\centering\arraybackslash}p{2.5cm}|>{\centering\arraybackslash}p{2.5cm}|>{\centering\arraybackslash}p{2.5cm}|>{\centering\arraybackslash}p{2.5cm}}
		\centering $\Lambda$CDM      & Baseline & + LSS  & + S$H_0$ES & + S$H_0$ES + LSS \\ \hline\hline
		Pl.2018 low $\ell$ TT        & 23.2     & 23.1   & 22.3       & 22.3             \\ \hline
		Pl.2018 low $\ell$ EE        & 395.9    & 397.5  & 397.1      & 395.9            \\ \hline
		Pl.2018 high $\ell$ TTTEEE   & 2338.2   & 2338.8 & 2342.6     & 2341.7           \\ \hline
		Pl.2018 Lensing              & 8.8      & 8.8    & 9.4        & 9.4              \\ \hline
		BAO SDSS DR7 mgs             & 1        & 1.8    & 2.3        & 2.3              \\ \hline
		BAO 6dF 2011                 & 0.1      & 0      & 0          & 0                \\ \hline
		BAO SDSS DR12                & 5.2      & 3.4    & 3.5        & 3.5              \\ \hline
		SN Pantheon                  & 1035.2   & 1034.8 & 1034.7     & 1034.7           \\ \hline
		SH$0$ES 2021                 &          &        & 17.9       & 17.8             \\ \hline
		LSS                          &          & 4.6    &            & 3.1              \\ \hline\hline
		\centering $\Sigma\, \chi^2$ & 3807.6   & 3812.8 & 3829.8     & 3830.7           \\ \hline
	\end{tabular}
	\caption{Bestfit $\chi^2$ values for the different datasets used in the MCMC chains using the $\Lambda$CDM model.}
	\label{tab:chiSq-LCDM}
\end{table}

\begin{table}[!htbp]
	\centering
	\renewcommand{\arraystretch}{1.1}
	\setlength\tabcolsep{2pt}
	\fontsize{8}{11}\selectfont
	\begin{tabular}{>{\arraybackslash}p{3.5cm}|>{\centering\arraybackslash}p{2.5cm}|>{\centering\arraybackslash}p{2.5cm}|>{\centering\arraybackslash}p{2.5cm}|>{\centering\arraybackslash}p{2.5cm}}
		\centering NEDE $(\Omega_{\phi} = 0)$ & Baseline & + LSS  & + S$H_0$ES & + S$H_0$ES + LSS \\ \hline\hline
		Pl.2018 low $\ell$ TT                 & 21.8     & 22.9   & 20.9       & 20.6             \\ \hline
		Pl.2018 low $\ell$ EE                 & 396.3    & 396.0  & 396.2      & 396.5            \\ \hline
		Pl.2018 high $\ell$ TTTEEE            & 2336.9   & 2338.6 & 2337.4     & 2339.0           \\ \hline
		Pl.2018 Lensing                       & 9.1      & 9.2    & 9.9        & 9.9              \\ \hline
		BAO SDSS DR7 mgs                      & 1.6      & 1.5    & 1.8        & 2.4              \\ \hline
		BAO 6dF 2011                          & 0.0      & 0.0    & 0.0        & 0.0              \\ \hline
		BAO SDSS DR12                         & 3.7      & 3.7    & 3.5        & 3.7              \\ \hline
		SN Pantheon                           & 1034.8   & 1034.9 & 1034.8     & 1034.8           \\ \hline
		SH$0$ES 2021                          & --       & --     & 1.5        & 1.0              \\ \hline
		LSS                                   & --       & 4.9    & --         & 7.7              \\ \hline\hline
		\centering $\Sigma\, \chi^2$          & 3804.2   & 3811.7 & 3805.9     & 3815.6           \\ \hline
	\end{tabular}
	\caption{Bestfit $\chi^2$ values for the different datasets used in the MCMC chains using the NEDE implementation with $\Omega_\phi = 0$ from \cite{Niedermann:2020dwg}.}
	\label{tab:chiSq-NEDEomega0}
\end{table}

\begin{table}[!htbp]
	\centering
	\renewcommand{\arraystretch}{1.1}
	\setlength\tabcolsep{2pt}
	\fontsize{8}{11}\selectfont
	\begin{tabular}{>{\arraybackslash}p{3.5cm}|>{\centering\arraybackslash}p{2.5cm}|>{\centering\arraybackslash}p{2.5cm}|>{\centering\arraybackslash}p{2.5cm}|>{\centering\arraybackslash}p{2.5cm}}
		\centering NEDE $(\Omega_{\phi} > 0)$ & Baseline & + LSS   & + S$H_0$ES${}^\lozenge$ & + S$H_0$ES + LSS \\ \hline\hline
		Pl.2018 low $\ell$ TT                    & 21.09    & 21.40   & 20.8       & 21.28            \\ \hline
		Pl.2018 low $\ell$ EE                    & 396.32   & 396.12  & 396.2      & 396.11           \\ \hline
		Pl.2018 high $\ell$ TTTEEE               & 2336.86  & 2340.78 & 2337.4     & 2340.06          \\ \hline
		Pl.2018 Lensing                          & 9.60     & 9.18    & 9.9        & 9.72             \\ \hline
		BAO SDSS DR7 mgs                         & 1.78     & 1.65    & 1.8        & 1.62             \\ \hline
		BAO 6dF 2011                             & 0.00     & 0.00    & 0.0        & 0.00             \\ \hline
		BAO SDSS DR12                            & 3.52     & 3.67    & 3.5        & 3.72             \\ \hline
		SN Pantheon                              & 1034.77  & 1034.80 & 1034.8     & 1034.80          \\ \hline
		S$H_0$ES 2021                             & --       & --      & 1.5        & 1.31             \\ \hline
		LSS                                      & --       & 2.8     & --         & 1.93             \\ \hline\hline
		\centering $\Sigma\, \chi^2$             & 3803.9   & 3810.4  & 3805.9     & 3810.56          \\ \hline
	\end{tabular}\\[5pt]
	{\footnotesize ${}^\lozenge$ We quote the same result as for the $\Omega_\phi = 0$ case since no minimization output beat the result in Table~\ref{tab:chiSq-NEDEomega0}. }
	\caption{Bestfit $\chi^2$ values for the different datasets used in the MCMC chains using the NEDE implementation with $\Omega_\phi > 0$.}
	\label{tab:chiSq-NEDE}
\end{table}

\bibliographystyle{jhep}
\bibliography{references}

\end{document}